\documentclass[12pt]{article}

\ifx\pdfoutput\undefined
\usepackage[dvips,bookmarks]{hyperref}
\else
\usepackage{hyperref}
\fi
\hypersetup{colorlinks=false,bookmarksopen,bookmarksnumbered,citecolor=blue,
   pdfstartview=FitH}

\usepackage[dvips]{graphicx}
\usepackage{latexsym}

\usepackage{amssymb,amsfonts,amsmath}
\usepackage{graphicx} 
\usepackage{indentfirst}

 \usepackage{bbm}

\parskip=\medskipamount

\newcommand {\cF}{{\cal F}}

\newcommand {\cH}{{\cal H}}

\newcommand {\cL}{{\cal L}}
\newcommand {\cM}{{\cal M}}
\newcommand {\cN}{{\cal N}}
\newcommand {\cO}{{\cal O}}

\def\a{\alpha}
\def \bi{\bibitem}

\def\b{\beta}
\def\c{\chi}
\def\d{\delta}

\def\g{\gamma}
\def\G{\Gamma}

\def\j{\psi}

\def\o{\omega}
\def\p{\pi}
\def\q{\theta}

\def\s{\sigma}

\def\x{\xi}
\def\z{\zeta}
\def\D{\Delta}
\def\F{\Phi}
\def\J{\Psi}
\def\L{\Lambda}
\def\O{\Omega}
\def\P{\Pi}

\def\S{\Sigma}
\def\U{\Upsilon}
\def\X{\Xi}

\def\rd{{\rm d}}
\def\ri{{\rm i}}

\newcommand{\ad}{{\dot{\alpha}}}                           %new
\newcommand{\bd}{{\dot{\beta}}}                            %new
\newcommand{\ve}{\varepsilon}                            %new
\newcommand{\pa}{\partial}                           %new
\newcommand{\hf}{\frac12}
\newcommand{\be}{\begin{equation}}
\newcommand{\ee}{\end{equation}}
\newcommand{\bea}{\begin{eqnarray}}
\newcommand{\eea}{\end{eqnarray}}
\newcommand{\non}{\nonumber}
\newcommand{\1}{\underline{1}}
\newcommand{\2}{\underline{2}}

\def\dt#1{{\buildrel {\hbox{\LARGE .}} \over {#1}}}    % dot-over for sp/sb

\newcommand{\bm}[1]{\mbox{\boldmath$#1$}}

\def\double #1{#1{\hbox{\kern-2pt $#1$}}}

\begin{document}
%%%%%%%%%%%%%%%%
%%%%%%%%%%%%%%%%
\begin{titlepage}
\begin{flushright}
UUITP-23/10\\
June, 2010\\
\end{flushright}
\vspace{5mm}

\begin{center}
{\Large \bf New   supersymmetric $\s$-model duality 
}\\ 
\end{center}

\begin{center}

{\bf
Sergei M. Kuzenko\footnote{kuzenko@cyllene.uwa.edu.au}${}^{a}$,
Ulf Lindstr\"om\footnote{ulf.lindstrom@fysast.uu.se}${}^{b}$
and Rikard von Unge\footnote{unge@physics.muni.cz}${}^{c}$ 
} \\
\vspace{5mm}

\footnotesize{
${}^{a}${\it School of Physics M013, The University of Western Australia\\
35 Stirling Highway, Crawley W.A. 6009, Australia}}  
~\\
\vspace{2mm}

\footnotesize{
${}^{b}${\it Theoretical Physics, Department of Physics and Astronomy,
Uppsala University \\ 
Box 803, SE-751 08 Uppsala, Sweden}
}
\\
\vspace{2mm}

\footnotesize{
${}^c${\it Institute for Theoretical Physics, Masaryk University, \\
61137 Brno, Czech Republic } }

\end{center}
\vspace{5mm}

\begin{abstract}
\baselineskip=14pt
We study dualities in off-shell 4D $\cN=2$ supersymmetric $\s$-models, 
using the projective superspace approach.
These include {\it (i)} duality between the real $\cO (2n)$ and  polar multiplets;
and {\it (ii)} polar-polar duality. We demonstrate that the dual of any superconformal 
$\s$-model is superconformal. Since $\cN=2$ superconformal 
$\s$-models (for which target spaces are hyperk\"ahler cones)
formulated in terms of polar multiplets are naturally associated with K\"ahler cones
(which are target spaces for $\cN=1$ superconformal $\s$-models), polar-polar duality
generates a transformation between different K\"ahler cones.
In the non-superconformal case, we study implications of polar-polar duality
for the $\s$-model formulation in terms of $\cN=1$ chiral superfields.
In particular, we find the relation between the original hyperk\"ahler potential and its dual.
As an application of polar-polar duality, we study self-dual models.
\end{abstract}
\vspace{1cm}

\vfill
\end{titlepage}

\newpage
\renewcommand{\thefootnote}{\arabic{footnote}}
\setcounter{footnote}{0}

\tableofcontents{}
\vspace{1cm}
\bigskip\hrule

\section{Introduction}
\setcounter{equation}{0}

Dualities in supersymmetric theories have a long history. 
In four-dimensional $\s$-models the duality between scalar and tensor multiplets, e.g., 
was  discussed in $\cN=1$ superspace both for $\cN=1$ and $\cN=2$ models already 
in \cite{Lindstrom:1983rt}. Here we shall be interested in $\cN=2$ supersymmetric $\s$-models 
and their dualities. These are best described in projective superspace \cite{KLR,LR} where 
the $\cN=2$ supersymmetry is manifest.\footnote{See \cite{Rosly,GIKOS} for alternative 
approaches.  General $\cN=2$ supersymmetric $\s$-models in harmonic superspace
and their dualities were studied in \cite{GIO}. Such $\s$-models do not possess 
a natural decomposition in terms of standard $\cN=1$ superfields, 
a property that is desirable for various applications.  
The existence of such a decomposition is one of the powerful inborn features 
of $\cN=2$ multiplets in projective superspace.}
There are several types of dualities (obtained by applying generalized
Legendre transformations)  for off-shell multiplets in projective  superspace.
These include  {\it (i)}  duality between so-called real $\cO (2n)$ and  polar multiplets
which was  introduced in \cite{LR} (see also \cite{G-RRWLvU});
and {\it (ii)} the duality between polar multiplets  which was 
introduced in \cite{GK} and also studied in \cite{LR10}.
Of particular interest to us here is the latter, polar-polar duality. 

We pay special attention to off-shell $\cN=2$ superconformal  $\s$-models
formulated in terms of projective superconformal multiplets \cite{K-hyper}.
As always, superconformal invariance is of interest in itself.
What is more important, the $\s$-models under consideration 
can be coupled to $\cN=2$ conformal supergravity \cite{KLRT-M}.

Superconformal $\s$-model dynamics turns out to require interesting target space geometry.
In the component approach, general $\cN=2$ superconformal $\s$-models were studied in 
\cite{deWKV1,deWKV2,deWRV} (see also \cite{SezginT}). Their target spaces are 
hyperk\"ahler spaces possessing a homothetic conformal Killing vector which is  the gradient of a function, 
and hence an isometric action of SU(2) rotating the complex structures. 
Such spaces are known as ``hyperk\"ahler cones'' \cite{deWRV} and they are intimately related to 
quaternion K\"ahler manifolds which are target spaces for $\cN=2$ locally supersymmetric $\s$-models
\cite{BW}. Specifically, there exists a one-to-one correspondence  \cite{Swann}
(see also \cite{Galicki}) between $4n$-dimensional quaternion K\"ahler manifolds
and $4(n+1)$-dimensional hyperk\"ahler cones. 

In the projective superspace approach, general off-shell $\cN=2$ superconformal $\s$-models were studied 
in \cite{K-hyper,KLvU,K-duality}, while the superconformal couplings of $\cN=2$ tensor multiplets
had appeared already in \cite{KLR} without a discussion of the conformal properties. 
The superconformal couplings of $\cN=2$ tensor multiplets were systematically 
discussed in the component approach in \cite{deWRV}.

General off-shell  $\cN=2$ superconformal $\s$-models are associated with K\"ahler cones
\cite{K-hyper,KLvU,K-duality}.
If the dimension of the $\s$-model hyperk\"ahler target space is $4n$, 
then the associated K\"ahler cone (following the terminology of \cite{GR,BCSS})
has dimension $2n$. As defined in \cite{GR,BCSS},  a K\"ahler cone $\cM$
is a K\"ahler space possessing a 
homothetic conformal Killing vector which is  the gradient of a function, and therefore holomorphic.
If $(\c^I , {\bar \c}^{\bar J})$ are the components of the homothetic conformal Killing vector, 
and $g_{I \bar J}$ is the K\"ahler metric, then
\begin{subequations}
\bea
\nabla_I \c^J &=& \d_I^J~, \qquad {\bar \nabla}_{\bar I} \c^J ={\bar \pa}_{\bar I} \c^J =0
 \label{1.1a} \\
 \c_I := g_{I\bar J}  {\bar \c}^{\bar J}&=& \pa_I K~, \qquad g_{I\bar J} 
%= {\bar \pa}_{\bar J} \pa_I K
=\pa_I {\bar \pa}_{\bar J} K
~, \label{1.1b}
\eea
\end{subequations}
where $K$ can be chosen to be 
\bea
 K =g_{I\bar J} \c^I  {\bar \c}^{\bar J}~.
 \eea
 We can choose local complex coordinates, $\F^I$, on $\cM$ in such a way that $\c^I =\F^I$.
%Its K\"ahler potential, $K(\F^I, {\bar \F}^{\bar J}) $,
% can be chosen to obey the homogeneity condition 
Then $K(\F^I, {\bar \F}^{\bar J}) $ obeys the following homogeneity condition:
\bea
\F^I \frac{\pa}{\pa \F^I} K(\F, \bar \F) =  K( \F,   \bar \F)~.
\label{Kkahler2}
\eea
Any K\"ahler cone is a cone \cite{GR}.
If $\cM$ in the above discussion is hyprek\"ahler, 
%similar to that given in for the 
it is called a hyperk\"ahler cone \cite{deWRV}. For the general properties of hyperk\"ahler cones, 
see \cite{deWRV,GR}. As shown in \cite{K-hyper,KLvU,K-duality}, the target spaces for 
general off-shell  $\cN=2$ superconformal $\s$-models,  which are hyperk\"ahler cones, are 
locally cotangent bundles over K\"ahler cones.

As is seen from (\ref{1.1b}), the function $K( \F,   \bar \F)$ can be identified with
the K\"ahler potential of $\cM$.
K\"ahler cones are  target spaces for $\cN=1$ superconformal $\s$-model, see, e.g.,
\cite{K-duality} for a detailed discussion.
The relationship between the hyperk\"ahler potential in the target space of a
$\cN=2$ superconformal $\s$-model and the associated K\"ahler cone was elaborated
in some detail in \cite{K-duality}.

At the level of $\cN=2$ superfields polar-polar duality amounts to a particular 
diffeomorphism \cite{LR10}.\footnote{More specifically, locally it is a symplectomorphism amounting to  
a change of polarization for the Darboux coordinates that describe the (2,0) holomorphic symplectic form
of the hyperk\"ahler manifold as it fibers the ${\mathbb C}P^1$ of complex structures.} 
Here we shall see that the $\cN=1$ interpretation is considerably more interesting. 
 It turns out that  polar-polar duality exchanges one K\"ahler cone with a different (dual) cone. 
 Since any $\cN=1$ superconformal $\s$-model has a K\"ahler cone as its target space, we may interpret 
 polar-polar duality as a transformation in the set of $\cN=1$ superconformal $\s$-models.

We further discuss the interpretation of polar-polar duality for the non-superconformal 
$\s$-models in terms of physical $\cN=1$ fields and show that it defines a transformation 
of certain $n$-dimensional K\"ahler spaces to other $n$-dimensional  K\"ahler spaces.

${}$Finally, as an important application, polar-polar duality allows us to introduce the family of self-dual models.

The paper is organized as follows: In section 2 we recapitulate some salient features of 
projective superspace and the definition  of superconformal projective mutiplets.
Section 3 starts our duality discussion by  providing a manifestly $\cN=2$ supersymmetric 
(and, where appropriate, superconformal) description 
of $\cO(2n)$/polar and  polar/polar dualities 
%at the level of 
in terms of $\cN=2$ projective superfields. In section 4 we examine these dualities when reduced 
to $\cN=1$ superspace.  One of the main results obtained in sections 3 and 4 is the proof of the fact  
that the dual of any $\cN=2$ superconformal $\s$-model is superconformal. 
Our analysis is deepened and carried out in more detail in section 5 for models with one 
polar multiplet. This section also contains several important examples.  
In section 6 we extend the analysis of the previous section to models containing
a set of $n$ polar multiplets, 
again giving several examples. Section 7 contains a discussion of the intriguing possibility of 
self-dual models in the present setting, while section 8 contains a few concluding comments.
We have collected some relevant features of superconformal Killing vectors in Appendix A. Finally,
in Appendix B we discuss properties of the tensor multiplet formulation for 
$\sigma$-models with  U(1)$\times$U(1) symmetry (\ref{5.43}). 
Review material is collated in section 2 and Appendix A.

\section{Superconformal projective multiplets}
\setcounter{equation}{0}
\label{sec2}

We start from the algebra of $\cN=2$ spinor covariant derivatives\footnote{Internal indices
take two values,  $i, j =\1, \2$. We use underlined symbols to avoid notational 
confusion (say, between $D^{\2}$ and $D^2 = D \cdot D$).}
\bea
\{D^i_{\a} \, , \, D^j_{ \b} \} = 0~,
\quad 
\{{\bar D}_{\dt \a }^i \, , \, {\bar D}_{\dt  \b }^j \} = 0~,  \quad 
\{D^i_{\a} \, , \, \bar D_{ \dt \b }^j \} = 2{\rm i} \, \ve^{i j}\,
(\s^m )_{\a \dt \b} \,\pa_m ~.
\label{5.2}
\eea
These  relations  encode an important structure that can be uncovered
by introducing  an auxiliary isotwistor 
$ v^i \in {\mathbb C}^2 \setminus \{0\} $
%=: \mathbb{C}^* $ 
and defining the following operators: 
${\mathfrak D}_{ \a} :=v_i \,{ D}^i_{ \a}$ and ${\bar {\mathfrak D}}_{\dt \a} := v_i \,{\bar { D}}^i_{\dt \a}$.
Then, the anti-commutation  relations (\ref{5.2}) imply that 
\bea
\{ {\mathfrak D}_{ \a} , {\mathfrak D}_{ \b} \} = \{{\mathfrak D}_{ \a} , {\bar {\mathfrak D}}_{\dt \b }\} 
=\{ {\bar {\mathfrak D}}_{\dt \a} , {\bar {\mathfrak D}}_{\dt \b} \}=0~.
\eea
These  identities constitute  the integrability conditions for  existence of certain constrained 
$\cN=2$ superfields that live in ${\mathbb R}^{4|8}\times {\mathbb C}P^1$ and are
annihilated by  ${\mathfrak D}_{ \a}$ and ${\bar {\mathfrak D}}_{\dt \a} $.

${}$Following \cite{K-hyper}, a superconformal projective multiplet of weight $n$,
$Q^{(n)}(z,v)$, is a superfield that 
lives on  ${\mathbb R}^{4|8}$, 
is holomorphic with respect to 
 $v^i $ on an open domain of 
${\mathbb C}P^1 $, 
and is characterized by the following conditions:\\
($a$) it obeys the analyticity constraints 
\be
{\mathfrak D}_{\a} Q^{(n)} ={\bar {\mathfrak D}}_{\dt \a} Q^{(n)} =0~;
\label{ana}
\ee  
($b$) it is  a homogeneous function of $v^i$ 
of degree $n$, that is  
\be
Q^{(n)}(z,c\,v)\,=\,c^n\,Q^{(n)}(z,v)~, \qquad c\in {\mathbb C} \setminus \{0\} \equiv \mathbb{C}^*~;
\label{weight}
\ee
($c$) it obeys the following $\cN=2$ superconformal transformation law:
\be
\d Q^{(n)} = - \Big(  \x -  \frac{\L^{(2)}}{(v,u)} \,u^i \frac{\pa}{\pa v^i}  \Big) \, Q^{(n)} 
-n \,\S \, Q^{(n)} ~.
\label{promult1}
\ee
Here $\x =\x^A (z)D_A$ is a $\cN=2$ superconformal Killing vector, 
\bea
\L^{(2)} := \L_{ij} (z)v^iv^j~, \qquad \S =  \frac{ \L_{ij} (z)v^i u^j   }{(v,u)} \,  + {\s} (z)
+\bar{ {\s} }(z)~, \qquad 
\eea
and $ \L_{ij} (z)$ and $\s (z)$ are related to  $\x$ as in eqs. (\ref{4DmasterN=2})--(\ref{lambdaN=2}). 
In the transformation law (\ref{promult1}), 
$u_i$ denotes  a {\it fixed} isotwistor chosen to be arbitrary modulo 
the condition $(v,u):= v^i u_i \neq 0$. Both $  Q^{(n)} $ and $\d Q^{(n)} $ are independent of $u_i$.
The parameters $\S$ and $\L^{(2)}$ obey the identities:
\be
{\mathfrak D}_\a \, \L^{(2)} = {\bar {\mathfrak D}}_{\dt \a} \, \L^{(2)}=0~, \qquad 
{\mathfrak D}_\a \, \S = {\bar {\mathfrak D}}_{\dt \a} \, \S=0~, \qquad 
u^i\frac{\pa}{\pa v^i}  \S =  \frac{\L^{(2)}}{(v,u)}~.
\label{Sigma}
\ee
 
Given a  superconformal  weight-$n$ multiplet $ Q^{(n)} (v^i)$, 
its {\it smile conjugate},\footnote{The   smile conjugation is the real structure
pioneered by Rosly \cite{Rosly} and re-discovered in \cite{GIKOS,KLR,HitchinKLR}.}
$ \breve{Q}^{(n)} (v^i)$, is defined by 
\bea
 Q^{(n)}(v^i) \longrightarrow  {\bar Q}^{(n)} ({\bar v}_i) 
  \longrightarrow  {\bar Q}^{(n)} \big({\bar v}_i \to -v_i  \big) =:\breve{Q}^{(n)}(v^i)~,
\label{smile-iso}
\eea
with ${\bar Q}^{(n)} ({\bar v}_i)  :=\overline{ Q^{(n)}(v^i )}$
the complex conjugate of  $ Q^{(n)} (v^i)$, and ${\bar v}_i$ the complex conjugate of 
$v^i$. One can show that $ \breve{Q}^{(n)} (v)$ is a superconformal  weight-$n$ multiplet,
unlike the complex conjugate of $Q^{(n)}(v) $.
One can also check that 
\bea
\breve{ \breve{Q}}^{(n)}(v) =(-1)^n {Q}^{(n)}(v)~.
\label{smile-iso2}
\eea
Therefore, if  $n$ is even, one can define real isotwistor superfields, 
 $\breve{Q}^{(2m)}(v) = {Q}^{(2m)}(v )$.

Our next goal is to understand how to engineer $\cN=2$ superconformal field theories 
described by superconformal projective multiplets.  
Let $\cL^{(2)}$ be  a  real superconformal  weight-2  multiplet, which is constructed in terms
of the dynamical superfields. Associated with $\cL^{(2)}$ is the superconformal action:
\bea
S:=- \frac{1}{2\p} \oint_{\g}  { v_i {\rm d} v^i }
\int {\rm d}^4x \, \D^{(-4)} \cL^{(2)}  \Big|_{\q={\bar \q} =0}~. 
\label{PAP}
\eea
Here $\g$ denotes a closed contour  in  ${\mathbb C}P^1$, $v^i(t)$,
parametrized by an evolution parameter $t$.
The action makes use of the following fourth-order differential operator:
\bea
\D^{(-4)} := \frac{1}{16} \nabla^\a \nabla_\a {\bar \nabla}_{\dt \b}  {\bar \nabla}^{\dt \b} ~, \qquad
\nabla_\a := \frac{1 }{ (v,u)} 
{ u_i} D^i_\a 
~, \quad 
{\bar \nabla}_{\dt \b} := \frac{1 }{(v,u)} 
u_i{\bar D}^i_{\dt \b} ~.~~
\eea
Here $u_i$ is defined below eq. (\ref{Sigma}),
and it is kept  fixed along the integration contour.
The action can be shown to be invariant under arbitrary  infinitesimal $\cN=2$ superconformal transformations
\cite{K-hyper}.
 
An important property of the action (\ref{PAP}) is its
invariance under  {projective transformations} of the form:
\be
\Big(u_i \,,\,v_i (t)\Big)~\to~\Big(u'_i \,,\,v'_i (t)\Big)=\Big(u_i\,,\, v_i(t) \Big)\,R~,~~~~~~R\,=\,
\left(\begin{array}{cc}a(t)~&0\\ b(t)~&c(t) \end{array}\right)\,\in\,{\rm GL(2,\mathbb{C})}~,
\label{projectiveGaugeVar}
\ee
where $t$ is the evolution parameter along the contour, and 
the matrix elements $a(t)$ and $b(t)$ obey the first-order equations:
\bea
{\dt a} = b \frac{({\dt v},v)}{(v,u)}~, \qquad 
{\dt b} = -b \frac{({\dt v},u)}{(v,u)}~, 
\label{2.11}
\eea
with $ {\dt \j}$ denoting the derivative of a function $\j(t)$ with respect to $t$.
Equations (\ref{2.11}) guarantee that 
the transformed isotwistor $u'_i$  is $t$-independent.  
This invariance allows one to make $u_i$ arbitrary modulo the constraint 
$(v,u)\neq 0$, and therefore the action is independent of $u_i$, that is 
$( \pa / \pa u_i ){ S}=0$.

Let  $\x_{\rm K}$ be a  superconformal Killing vector obeying the conditions
\bea
\L_{ij} (z) = \s (z) =0~,
\eea
with $\L_{ij} (z) $ and $  \s (z)$  defined in eqs. (\ref{lambdaN=2}) and (\ref{lor,weylN=2}), respectively.
It is called a $\cN=2$ Killing vector, for 
the set of all such vectors can be seen to form a superalgebra isomorphic to the $\cN=2$ super-Poincar\'e algebra. 
In the super-Poincar\'e case, the transformation law (\ref{promult1}) reduces to 
the universal (weight-independent) form:
\bea
\d Q^{(n)} = -  \x_{\rm K} \, Q^{(n)} ~.
\label{promult2}
\eea
If we are interested in general $\cN=2$ supersymmetric (i.e. super-Poincar\'e invariant)  theories, 
not necessarily superconformal ones, 
projective multiplets should be defined by the relations  (\ref{ana}), (\ref{weight}) and (\ref{promult2}).

Suppose we wish to construct  an off-shell  $\cN=2$ superconformal theory described by a given set 
of  superconformal projective multiplets ${\mathfrak P}_{A}^{(n_{A})} (z,v)$. 
Then, the corresponding Lagrangian
must be an  algebraic function of  the dynamical superfields, 
\bea
\cL^{(2)}_{\text{s-conformal}} = \cL ({\mathfrak P}_{A}^{(n_{A})})~,
\eea
and possess no  explicit dependence on the isotwistor $v^i$. Imposing the homogeneity condition
\bea
 \cL \big( c^{n_A}\, {\mathfrak P}_{A}^{(n_{A})}   \big) 
  = c^2 \,  \cL \big({\mathfrak P}_{A}^{(n_{A})}  \big) ~, \qquad c\in \mathbb{C}^*~.
\eea  
guarantees that $\cL^{(2)}_{\text{s-conformal}} $ is a superconformal
weight-two projective multiplet.
  
In the more general case of super-Poincar\'e invariant theories,   
the Lagrangian may depend explicitly on the isotwistor $v^i$,
\bea
\cL^{(2)}_{\text{s-Poincar\'e}} = \cL^{(2)} ({\mathfrak P}_{A}^{(n_{A})}; v)~,
\eea
and must obey the homogeneity condition 
\bea
 \cL \big( c^{n_A}\, {\mathfrak P}_{A}^{(n_{A})}; c \,v   \big) 
  = c^2 \,  \cL \big({\mathfrak P}_{A}^{(n_{A})} ;v \big) ~, \qquad c\in \mathbb{C}^*
\eea  
It is easy to show that the action (\ref{PAP}) generated by $\cL^{(2)}_{\text{s-Poincar\'e}} $
is $\cN=2$ supersymmetric. In the super-Poincar\'e case, the action (\ref{PAP}) can be shown to be 
equivalent to that proposed originally in \cite{KLR}.

Without loss of generality,  we can assume  that the integration contour $\g$
in (\ref{PAP}) does not pass through the ``{north pole}'' $v^{i}_{\rm north} \sim (0,1)$ of ${\mathbb C}P^1$.
It is then useful to introduce a complex 
(inhomogeneous) coordinate $\z$ in the north chart, $\mathbb C$, 
of ${\mathbb C}P^1 ={\mathbb C} \cup \{\infty \}$: 
\bea 
 v^i = v^{\1} \,(1, \z) ~,\qquad \z:=\frac{v^{\2}}{v^{\1}} ~,\qquad\quad 
{ i=\1 ,\2}
\label{Zeta}
\eea
and define projective multiplets in this chart.
Given a weight-$n$ projective superfield $ Q^{(n)}(z,v)$, we can associate with it 
a new object $ Q^{[n]}(z,\z )$ defined as 
\bea
Q^{(n)}(z,v)~\longrightarrow ~ Q^{[n]}(z,\z) \propto Q^{(n)}(z,v)~, \qquad 
\frac{\pa}{\pa \bar \z} Q^{[n]} =0 ~.
\eea
The explicit form of  $ Q^{[n]}(z,\z) $ depends on the multiplet under consideration, 
and will be specified below.
In terms of  $ Q^{[n]}(z,\z) $, the analyticity constraints (\ref{ana})
take the form: 
\bea
{ D^{\2}_{\a}}
Q^{[n]}(\z)=\z\,{ D^{\1}_{\a}}Q^{[n]}(\z)~, \qquad 
{ {\bar D}_{  {\dt \a}\, \2}}Q^{[n]}(\z)=-\frac{1}{\z}\,
{ {\bar D}_{ {\dt \a}\, \1}}Q^{[n]}(\z)~.
\label{ancon}
\eea
The   $Q^{[n]}(z,\z) $ can be represented by a Laurent series
\bea
Q^{[n]}(z,\z) = \sum Q_k (z) \z^k~,
\label{seriess}
\eea
with $Q_k(z)$ some {ordinary $\cN=2$ superfields}. In accordance with (\ref{smile-iso}),
the smile conjugate of ${Q}^{[n]} (z,\z) $ is defined as follows:
\bea
\breve{Q}^{[n]} (z, \z) :=\sum (-1)^k{\bar Q}_{k} (z) { \z^{-k}}~.
\eea
Unlike eq. (\ref{smile-iso2}), we now have 
\bea
\breve{\breve{Q}}{}^{[n]} (\z) ={Q}^{[n]} (\z)~.
\eea
A real projective superfield is characterized by the properties:
\bea
 \breve{Q}^{[n]} (z,\z) =Q^{[n]}(z, \z) =
{ \sum Q_k (z) \z^k}~, \qquad 
{\bar Q}_k = (-1)^k Q_{-k}.
\label{realPS}
\eea

When switching from $ Q^{(n)}(v)$ to $Q^{[n]}(\z )$, the information about 
the degree of homogeneity, $n$,  remains encoded only in the superconformal 
transformation law, eq. (\ref{promult1}). In the super-Poincar\'e case, 
the superscript $[n]$ becomes redundant and is usually omitted.

We conclude this introductory section by listing those projective multiplets which are used 
for $\s$-model constructions.
Our first example is the so-called real $\cO(2n)$ multiplet\footnote{Here and below, 
we use the terminology introduced originally in \cite{G-RRWLvU} 
for non-superconformal projective multiplets.} 
\cite{KLR,KLT,LR}, $n=1,2\dots$, which 
 is described by a real weight-$2n$ projective superfield $\eta^{(2n)} (z,v) $ of the form:
\bea
\eta^{(2n)} (z,v) &=& \eta_{i_1 \dots i_{2n}}(z) \,v^{i_1} \dots v^{i_{2n}} 
=\breve{\eta}^{(2n)} (z,v) ~.
\eea
Here $\eta_{i_1 \dots i_{2n}}(z) $ are completely symmetric $\cN=2$ superfields
obeying the constraints
\bea
D_{\a (j} \eta_{i_1 \dots i_{2n} )} ={\bar D}_{\dt \a (j} \eta_{i_1 \dots i_{2n} )} =0~
\label{2.28}
\eea
which follow from (\ref{ana}).
It should be pointed out that the reality condition $\breve{\eta}^{(2n)}  = {\eta}^{(2n)} $ is equivalent to 
\bea
\overline{ \eta_{i_1 \dots i_{2n}} } &=& \eta^{i_1 \dots i_{2n}}
=\ve^{i_1 j_1} \cdots \ve^{i_{2n} j_{2n} } \eta_{j_1 \dots j_{2n}} ~.
\eea
Associated with  $\eta^{(2n)} (z,v) $ is the superfield $\eta^{[2n]}(z,\z) $ defined by 
\bea
\eta^{(2n)} (z,v) &=&\big({\rm i}\, v^{\1} v^{\2}\big)^n \eta^{[2n]}(z,\z) =
\big(v^{\1}\big)^{2n} \big({\rm i}\, \z\big)^n \eta^{[2n]}(z,\z)~,  \non \\
\eta^{[2n]}(z,\z) &=& 
\sum_{k=-n}^{n} \eta_k (z) \z^k~,
\qquad  {\bar \eta}_k = (-1)^k \eta_{-k} ~.
\label{o2n1}
\eea
The superfield $\eta^{[2n]}(z,\z)$  is real in the sense of (\ref{realPS}).

To describe charged hypermultiplets, one uses the so-called {\it arctic}
multiplet $\U^{(n)} (z, v) $ \cite{LR}, which is defined to be 
holomorphic  in the north chart
of ${\mathbb C}P^1$,
\bea
\U^{(n)} (z, v) &=&  (v^{\1})^n\, \U^{[n]} (z, \z) ~, \qquad 
\U^{ [n] } (z, \z) = \sum_{k=0}^{\infty} \U_k (z) \z^k~, 
\label{arctic1}
\eea
and  its smile-conjugate {\it antarctic} multiplet $\breve{\U}^{(n)} (z,v) $,
 \bea
\breve{\U}^{(n)} (z,v) &=& (v^{\1} \,\z \big)^{n}\, \breve{\U}^{[n]}(z,\z) ~, \qquad
\breve{\U}^{[n]}( z,\z) = \sum_{k=0}^{\infty}  {\bar \U}_k (z)\,
\frac{(-1)^k}{\z^k}~.~~~
\label{antarctic1}
\eea
The pair $\U^{[n]} ( \z)$ and $\breve{\U}^{[n]}(\z) $ constitute the so-called polar multiplet.
The components  $\U_k (z)$ in (\ref{arctic1})  are constrained $\cN=2$ superfields, 
in accordance with (\ref{ancon}).

To describe gauge superfields ($n=0$) and Lagrange multipliers for various duality transformations, 
one uses the so-called real {\it tropical} multiplet $U^{(2n)} (z,v) $ \cite{LR} defined by 
\bea
U^{(2n)} (z,v) &=&\big({\rm i}\, v^{\1} v^{\2}\big)^n U^{[2n]}(z,\z) =
\big(v^{\1}\big)^{2n} \big({\rm i}\, \z\big)^n U^{[2n]}(z,\z)~,  \non \\
U^{[2n]}(z,\z) &=& 
\sum_{k=-\infty}^{\infty} U_k  (z)\z^k~,
\qquad  {\bar U}_k = (-1)^k U_{-k} ~.
\label{2n-tropica1}
\eea
The superfield $U^{[2n]}(z,\z)$  is real  in the sense of (\ref{realPS}).

The $\cN=2$ superconformal transformation laws of the superfields 
$\eta^{[2n]}(\z) $, $\U^{ [n] } ( \z) $, $\breve{\U}^{[n]}( \z) $ and $U^{[2n]}(\z) $ 
are given in \cite{K-hyper}.

\section{Formulation in $\cN=2$ superspace}
\setcounter{equation}{0}
\label{sec3}

The formalism presented in the previous section is convenient for the formulation 
of manifestly $\cN=2$ supersymmetric duality transformations.
The main purpose of this section is to show that the dual of any superconformal 
field theory is superconformal.

\subsection{Duality between the real $\cO (2n)$ and  polar multiplets} 
\label{sec3.1}
Consider an off-shell $\cN=2$  supersymmetric $\s$-model 
described by  an $\cO (2n)$-multiplet $\eta^{(2n)} (z,v)$ and some other projective 
multiplets  $\O_a^{(n_a)} (z,v)$. Let $\cL^{(2)} (\eta^{(2n)}, \O_a^{(n_a)} ; v)$
be the Lagrangian of the theory. Note  that, 
in general, $\cL^{(2)}$ may explicitly depend on the isotwistor $v^i$.
In the superconformal case, however, the Lagrangian must be $v$-independent, 
$\cL^{(2)} (\eta^{(2n)}, \O_{a}^{(n_{a})} )$.

The theory under consideration has a dual formulation given by a different Lagrangian 
$\cL_{\rm D}^{(2)} ( \X^{(2-2n)},  \breve{\X}^{(2-2n)}, \O_{a}^{(n_{a})} ; v)$, 
in which $\X^{(2-2n)}$ is an arctic multiplet, 
and $\breve{\X}^{(2-2n)}$ its smile conjugate antarctic multiplet.
The dual description is obtained by Legendre transformation.
One proceeds  by replacing the original system by an auxiliary  first-order formulation
with Lagrangian
\bea
\cL^{(2)}_{\text{first-order} }
= \cL^{(2)} (U^{(2n)}, \O_{a}^{(n_{a})} ; v) + U^{(2n)} \big( \X^{(2-2n)} + \breve{\X}^{(2-2n)} \big)~,~~~
\label{firsr-order-O2n}
\eea
where $U^{(2n)}$ is a  {\it real tropical multiplet}. This model is equivalent to the original one.
Indeed, varying the first-order action, $S_{\text{first-order}}$, with respect to 
$\X^{(2-2n)}$ and $\breve{\X}^{(2-2n)}$ proves to constrain $U^{(2n)}$ 
to become  a real $\cO(2n)$ multiplet,
\bea
\frac{\d }{\d \X^{(2-2n)} }S_{\text{first-order} }=0 \quad \Longrightarrow \quad
U^{(2n)} = \eta^{(2n)}~,
\label{3.2}
\eea
and then $S_{\text{first-order}}$ reduces 
to the original action. On the other hand, varying the first-order action with respect to $U^{(2n)}$ 
gives\footnote{Since $ \X^{(2-2n)}$ and  $U^{(2n)}$ are constrained $\cN=2$ superfields, 
the equations (\ref{3.2}) and (\ref{3.3}) are quite nontrivial. They can be derived using a formulation 
in terms of $\cN=1$ superfields, as was done in the original publications \cite{LR,G-RRWLvU}; 
see also subsection \ref{general} below.}
\bea
\frac{\pa}{\pa U^{(2n)} } \cL^{(2)} (U^{(2n)}, \O_{a}^{(n_{a})} ; v) +  \X^{(2-2n)} + \breve{\X}^{(2-2n)} =0~.
\label{3.3}
\eea
Suppose this equation allows us to uniquely express $U^{(2n)}$ 
as a function of the other variables, that is
$U^{(2n)} = {\bm U}^{ (2n)} (  \X^{(2-2n)} , \breve{\X}^{(2-2n)} ,\O_{a}^{(n_{a})}; v)$.
Then, the dual Lagrangian is defined by 
\bea
\cL_{\rm D}^{(2)} ( \X^{(2-2n)},  \breve{\X}^{(2-2n)}, \O_{a}^{(n_{a})} ; v)
&{}& \non \\
=\Big\{ \cL^{(2)} (U^{(2n)}, \O_{a}^{(n_{a})} ; v) &+& U^{(2n)} \big( \X^{(2-2n)} + \breve{\X}^{(2-2n)} \big) \Big\}
\Big|~,
\eea
where the vertical stroke on the right indicates that the variable $U^{(2n)} $ should be replaced by its on-shell value
$ {\bm U}^{ (2n)} (  \X^{(2-2n)} , \breve{\X}^{(2-2n)} ,\O_{a}^{(n_{a})}; v)$.

The duality transformation presented  is compatible with $\cN=2$ superconformal invariance.
Indeed, suppose the original model is superconformal, and hence its Lagrangian has no explicit 
$v$-dependence, $\cL^{(2)} = \cL^{(2)} (\eta^{(2n)}, \O_{a}^{(n_{a})} )$.
It leads to the  first-order Lagrangian
\bea
\cL_{\text{first-order} }=
\cL^{(2)} (U^{(2n)}, \O_{a}^{(n_{a})} ) + U^{(2n)} \big( \X^{(2-2n)} + \breve{\X}^{(2-2n)} \big)~,
\label{firsr-order-O2n-2}
\eea
which also has no explicit $v$-dependence, and therefore generates a superconformal  theory. 
Integrating out $U^{(2n)}$ does not generate any explicit $v$-dependence.
We conclude  that the dual Lagrangian is $v$-independent, 
$\cL_{\rm D}^{(2)} =\cL_{\rm D}^{(2)} ( \X^{(2-2n)},  \breve{\X}^{(2-2n)}, \O_{a}^{(n_{a})} )$, and therefore 
the dual theory is  $\cN=2$ superconformal.

\subsection{Polar-polar duality}
\label{Polar-polar duality I}
A different type of duality  can be defined in the case of 
 a nonlinear $\s$-model in which the dynamical variables include a polar multiplet 
realized in terms of an arctic superfield $\U^{(n)} (z,v)$  
and  its smile conjugate antarctic superfield $\breve{\U}^{(n)} (z,v)$.
Along with this polar multiplet,
the theory may also describe the dynamics of some other multiplets  $\O_{a}^{(n_{a})}(z,v)$.
We denote the corresponding Lagrangian by $\cL^{(2)} (\U^{(n)}, \breve{\U}^{(n)} ,\O_{a}^{(n_{a})} ; v)$.

The theory under consideration possesses an equivalent first-order formulation generated by
\bea
\cL^{(2)}_{\text{first-order} }
= \cL^{(2)} (W^{(n)}, \breve{W}^{(n)}, \O_{a}^{(n_{a})} ; v) + {\rm i}\, W^{(n)}  \X^{(2-n)} 
- {\rm i} \, \breve{W}^{(n)} \breve{\X}^{(2-n)} ~,~~~
\label{firsr-order-polar}
\eea
where $W^{(n)}$ is {\it complex tropical}, and $\X^{(2-n)}$  {\it arctic}. 
Let $S_{\text{first-order} }$ be the corresponding action.
Indeed, it will be shown in subsection
\ref{general} that 
the equation of motion for $ \X^{(2-n)} $ implies that $ W^{(n)}$ is a weight-$n$ arctic multiplet, 
\bea
\frac{\d }{\d \X^{(2-n)} }S_{\text{first-order} }=0 \quad \Longrightarrow \quad
W^{(n)} = \U^{(n)}~.
\label{3.7}
\eea
Then,  the action $S_{\text{first-order} }$ reduces to that generated 
by $\cL^{(2)} (\U^{(n)}, \breve{\U}^{(n)} ,\O_{a}^{(n_{a})}  ; v)$.
On the other hand,  the equations of motion for $ W^{(n)} $ and $\breve{W}^{(n)}$ are:
\begin{subequations}
\bea
\frac{\pa}{\pa W^{(n)} } \cL^{(2)} (W^{(n)}, \breve{W}^{(n)}, \O ; v) &+&{\rm i}\,   \X^{(2-n)} =0~,
\label{3.8a} \\
\frac{\pa}{\pa \breve{W}^{(n)} } \cL^{(2)} (W^{(n)}, \breve{W}^{(n)}, \O ; v) &-& 
{\rm i}\,  \breve{\X}^{(2-n)} =0~.
\label{3.8b}
\eea
\end{subequations}
Under rather  general assumptions, these algebraic equations  
can be used to express   $ W^{(n)} $ and $\breve{W}^{(n)}$ 
in terms of the other variables. This leads to the dual Lagrangian:
\bea
\cL^{(2)}_{\rm D} (\X^{(2-n)},  \breve{\X}^{(2-n)}  ,\O ; v) &=& \Big\{ 
\cL^{(2)} (W^{(n)}, \breve{W}^{(n)}, \O ; v) \non \\
&& \qquad \qquad + {\rm i}\, W^{(n)}  \X^{(2-n)} 
- {\rm i} \, \breve{W}^{(n)} \breve{\X}^{(2-n)} \Big\} \Big|~, 
\eea
where the vertical stroke on the right indicates that the variables  $ W^{(n)} $ and $\breve{W}^{(n)}$ 
should be replaced by their  on-shell values.
In the special case $n=1$, both the original and dual polar multiplets 
have the same weight.

\subsection{Polar-polar duality and superconformal $\s$-models}

We consider a system of interacting weight-one\footnote{To simplify the notation, 
in this subsection we denote $\U^+ \equiv \U^{(1)}$.}
 arctic  multiplets, 
$\U^{+ I} (z,v) $, and their smile-conjugates,
$ \breve{\U}^{+\bar I }(z,v)$, described by a Lagrangian of the form \cite{K-hyper}:
\bea
\cL^{(2)}  (\U^{+}, \breve{\U}^{+})= {\rm i} \, K (\U^{+}, \breve{\U}^{+})~,
\label{conformal-sm}
\eea
Here $K(\F^I, {\bar \F}^{\bar J}) $ is a real function
of $n$ complex variables $\F^I$, with $I=1,\dots, n$, 
which obeys the homogeneity condition (\ref{Kkahler2}).
The function  $K(\F^I, {\bar \F}^{\bar J}) $ can be interpreted as the K\"ahler potential 
of a {\it K\"ahler cone} (following the terminology of  \cite{GR,BCSS}).
Of course, this interpretation requires the K\"ahler metric 
$g_{I \bar J}:= K_{I \bar J} $ to be non-singular,
\bea
\det \,( K_{I \bar J} ) \neq 0~,
\label{3.11}
\eea
where we have used the standard  the notation:
\bea
K_{I_1 \dots I_p \,{\bar J}_1 \dots {\bar J}_q } &:=& 
\frac{\pa^{p+q} K}{\pa \F^{I_1} \dots \pa \F^{I_p}\, {\bar \F}^{{\bar J}_1} \dots {\bar \F}^{{\bar J}_q}}~.
\eea
The action 
\bea
S[\F, \bar \F] =  \int 
\rd^4 x\,{\rm d}^4\q
\, K(\Phi^{I},
 {\bar \Phi}{}^{\bar{J}})  ~,  \qquad {\bar D}_{\dt \a} \F^I=0~,
\label{nact4}
\eea
with  $K(\F^I, {\bar \F}^{\bar J}) $ obeying the homogeneity condition (\ref{Kkahler2}), 
defines the most general $\cN=1$ superconformal $\s$-model, see, e.g.,
\cite{K-duality}.

We are interested in the dual formulation for the theory (\ref{conformal-sm}) which is obtained 
by performing the polar-polar duality with respect to all the  multiplets:
\bea
\cL^{(2)}_{\rm D } ( \X^+, \breve{\X}{}^+ ) =  {\rm i} \,\Big\{ 
K(W^{+}, \breve{W}^{+})
+ W^{+I}  \X^{+}_I 
-\breve{W}^{+\bar I} \breve{\X}^{+}_{\bar  I} \Big\}\Big|~,
\label{conformal-Legendre}
\eea
where the vertical stroke on the right indicates that the complex tropical superfields 
$ W^{+I} $ and  their smile-conjugates $\breve{W}^{+\bar I} $ should be expressed in terms of 
weight-one arctic superfields $\X^+_I$ and their smile-conjugates 
$ \breve{\X}^{+}_{\bar  I} $ using the following equations of motion:
\begin{subequations}
\bea
 \frac{\pa}{\pa W^{+I} } K (W^{+}, \breve{W}^{+}) &+&  \X^{+}_I =0~, 
 \label{3.13a}\\
 \frac{\pa}{\pa \breve{W}^{+\bar I} } K (W^{+}, \breve{W}^{+}) &-&  \breve{\X}^{+}_{\bar I} =0~.
\label{3.13b}
\eea
\end{subequations}
This requires the K\"ahler potential $K(\F^I, {\bar \F}^{\bar J}) $ to obey the condition 
\bea
\det  \left(
\begin{array}{c  c}
 K_{IJ }    ~& ~  K_{I \bar J}   \\
K_{\bar I J} 
& K_{\bar I \bar J}   \\
\end{array}
\right) \neq 0~.
\label{Hessian1}
\eea

Making use of the equations (\ref{3.13a}) and (\ref{3.13b}), 
in conjunction with  (\ref{Kkahler2}) and the standard properties 
of the Legendre transformation, one can show that the dual Lagrangian (\ref{conformal-Legendre})
obeys the homogeneity condition:
\bea
\X^+_I \frac{\pa}{\pa \X^{+}_I } \cL^{(2)}_{\rm D } ( \X^+, \breve{\X}{}^+ ) = 
\cL^{(2)}_{\rm D } ( \X^+, \breve{\X}{}^+ ) ~.
\eea
As a result, we can represent
\bea
\cL^{(2)}_{\rm D}(\X^{+}, \breve{\X}^{+}) = {\rm i} \, K_{\rm D}(\X^{+}, \breve{\X}^{+})~,
\label{conformal-sm-dual}
\eea
where $K_{\rm D} (\J_I, {\bar \J}_{\bar J}) $ is a real analytic function
of $n$ complex variables $\J_I$, with $I=1,\dots, n$, 
which obeys the homogeneity condition
\bea
\J_I \frac{\pa}{\pa \J_I} K_{\rm D}(\J, \bar \J) =  K_{\rm D}( \J,   \bar \J)~.
\label{Kkahler2-dual}
\eea
This function can be interpreted to be the K\"ahler potential of a K\"ahler cone.
For such an interpretation to be consistent, 
the corresponding K\"ahler metric $g_{\rm D}{}^{I \bar J}:= K_{\rm D}{}^{I \bar J} $
should be nonsingular,
\bea
\det \,( K_{\rm D}{}^{I \bar J} ) \neq 0~,
\eea
This indeed follows from eqs. (\ref{3.11}) and (\ref{Hessian1}) of which the latter implies
\bea
\det  \left(
\begin{array}{c  c}
 K_{\rm D}{}^{IJ }    ~& ~  K_{\rm D}{}^{I \bar J}   \\
K_{\rm D}{}^{\bar I J} 
& K_{\rm D}{}^{\bar I \bar J}   \\
\end{array}
\right) \neq 0~.
\eea

We conclude that the $\cN=2$ {\it polar-polar  duality} transformation induces a transformation 
in the family  of $\cN=1$ superconformal $\s$-models. Specifically, the $\s$-model 
(\ref{nact4}) turns into
\bea
S_{\rm D}[\J, \bar \J] =  \int 
\rd^4 x\,{\rm d}^4\q
\, K_{\rm D}(\J_{I},
 {\bar \J}{}_{\bar{J}})  ~, \qquad {\bar D}_{\dt \a} \J_I=0~.
\label{nact5}
\eea

It should be emphasized that the above conclusions hold if the duality transformation
is applied to all the polar multiplets in the superconformal $\s$-model 
(\ref{conformal-sm}) and (\ref{Kkahler2}). Had we dualized some of the polar multiplets, 
we would have ended up with a dual formulation in which the Lagrangian obeys 
a different homogeneity condition.
Specifically, let us split the original set of arctic  multiplets, $\U^I$, into two subsets
$\U^I =(\U^i, \U^a)$, and apply the polar-polar duality to the first subset.
Then, we generate a dual Lagrangian
\bea
\cL^{(2)}_{\rm D}(\X^{+}_i , \U^{+a}, \breve{\X}^{+}_{\bar i}, \breve{\U}^{+\bar a}) 
\non
\eea
obeying the homogeneity condition 
\bea
\Big( \X^+_i \frac{\pa}{\pa \X^{+}_i }
+  \breve{\U}^{+\bar a}  \frac{\pa} {\pa  \breve{\U}^{+\bar a}} \Big)
 \cL^{(2)}_{\rm D}(\X^{+} , \U^{+}, \breve{\X}^{+}, \breve{\U}^{+ }) 
 =  \cL^{(2)}_{\rm D}(\X^{+} , \U^{+}, \breve{\X}^{+}, \breve{\U}^{+ }) ~.
\eea

\section{Formulation in $\cN=1$ superspace}
\setcounter{equation}{0}

${}$From the point of view of  various applications, one of the powerful properties of  projective 
multiplets, $Q^{[n]}(z,\z)$, is that they admit a simple decomposition in terms of standard $\cN=1$ 
superfields. This follows, in particular, from the analyticity constraints (\ref{ancon}) 
which can be interpreted as follows. For the component $\cN=2$ superfields $Q_k(z)$ of  $Q^{[n]}(z,\z)$
appearing in the series (\ref{seriess}), their dependence on $\q^\a_{\2}$ and ${\bar \q}^{\2}_{\dt \a}$
is uniquely determined, according to (\ref{ancon}), in terms of their dependence 
on the variables $\q^\a_{\1}=: \q^\a$ and ${\bar \q}^{\1}_{\dt \a}=: {\bar \q}_{\dt \a}$,
which can be identified with the 
Grassmann coordinates of $\cN=1$ superspace parametrized by 
$(x^m, \q^\a ,{\bar \q}_{\dt \a}$).\footnote{The $\cN=1$ spinor covariant derivatives 
are $D_\a :=D^{\1}_{\a}$ and $ { {\bar D}^{\dt \a}:= {\bar D}_{\1}^{\dt \a}}$.}
In other words, all information about the the projective
multiplet $Q^{[n]}(z,\z)$ is encoded in its $\cN=1$ projection
\bea
Q^{[n]}(x, \q_i,{\bar \q}^i,\z)\big|_{\q_{\2}={\bar \q}^{\2} =0}~.
\eea
What is the structure of  the $\cN=1$ superfields $Q_k\big|_{\q_{\2}={\bar \q}^{\2} =0}$
associated with the $\cN=2$ projective multiplet?
If the Laurent series (\ref{seriess}) terminates from below,
\bea
Q^{[n]}(z,\z) =
\sum_{p} Q_k (z) \z^k~,
\qquad  -\infty <p~,
\label{seriess2}
\eea
then the analyticity constraints (\ref{ancon}) imply that
the lowest components $Q_p $ and $Q_{p+1}$ are $\cN=1$ chiral and linear superfields, respectively.
\bea
{\bar D}_{\dt \a} Q_p &=&0~, 
\qquad {\bar D}^2 Q_{p+1} =0~. 
\eea
If the Laurent series (\ref{seriess}) terminates from above,
\bea
Q^{[n]}(z,\z) =
\sum^{q} Q_k (z) \z^k~,
\qquad  q< \infty~,
\label{seriess3}
\eea
then the analyticity constraints (\ref{ancon}) imply that
the highest components $Q_q $ and $Q_{q-1}$ are $\cN=1$ anti-chiral 
and anti-linear superfields, respectively.
\bea
D_\a Q_q =0~, \qquad D^2 Q_{q-1} =0~.
\eea
The other $\cN=1$ superfields $Q_k\big|_{\q_{\2}={\bar \q}^{\2} =0}$
in  (\ref{seriess}) 
turn out to be unconstrained, modulo possible reality conditions.

In the $\cN=2$ supersymmetric action (\ref{PAP}),  the Lagrangian $\cL^{(2)}$ is a projective multiplet,
and therefore it is fully determined by its  $\cN=1$ projection $\cL^{(2)}\big|_{\q_{\2}={\bar \q}^{\2} =0}$. 
Let us express the action (\ref{PAP}) in terms of this projection.
We recall that the integration contour $\g$
in (\ref{PAP}) is chosen to lie outside the ``{north pole}'' $v^{i}_{\rm north} \sim (0,1)$ of ${\mathbb C}P^1$,
which allows us to use the inhomogeneous complex coordinate, $ \z$, 
 defined by $ v^i = v^{\1} \,(1, \z)$.
Since the action is independent of $u_i$, the latter can be chosen to be   $ u_i =(1,0)$, 
such that $(v,u) = v^{\1}\neq 0$. 
We  represent the Lagrangian in the form: 
\bea
\cL^{(2)}(z,v)={\rm i} \,v^{\1}v^{\2}\cL(z,\z)
= {\rm i} (v^{\1})^2 \,\z\,{ \cL(z,\z)~, 
\qquad \breve{\cL} =\cL}~. 
\eea
It is important to remark that $\cL(z,\z)$ is a real projective superfield. 
Now, a short calculation (see, e.g. \cite{K10}) allows us to bring the action (\ref{PAP})  to the form: 
\bea
S 
&=& \frac{1 }{ 2\pi\ri }
 \oint_\g 
 \frac{\rd\z }{  \z}
\int\rd^4 x\,{\rm d}^4\q \,
\cL(z,\z)\Big|_{\q_{\2}={\bar \q}^{\2} =0}~.
\label{6.15}
\eea
Here the integration is carried out over the $\cN=1$ superspace.
The action is now formulated entirely in terms of $\cN=1$ superfields. 
At the same time, by construction, it is off-shell $\cN=2$ supersymmetric.

The main goal of this section is to reformulate the duality transformations, 
which we presented in  section \ref{sec3}.
in terms of $\cN=1$ superfields. In what follows, 
the symbol of $\cN=1$ projection in expressions like  (\ref{6.15}) is omitted.

\subsection{Duality between the real $\cO (2n)$ and  polar multiplets} 

We revisit the duality transformation between the real $\cO (2n)$ and  polar multiplets
considered in subsection \ref{sec3.1}.
Associated with the $\cO(2n)$ multiplet $\eta^{(2n)} (v) $ is the superfield $\eta^{[2n]}(\z) $ defined by 
eq. (\ref{o2n1}).
The two lowest components
in the expansion (\ref{o2n1}),
$\eta_{- n} $ and $\eta_{-n+1}$, are constrained $\cN=1$ superfields, chiral and linear, respectively, 
\bea
 {\bar D}_{\dt \a} \eta_{-n} &=&0~, 
\qquad {\bar D}^2 \eta_{-n+1} =0~. 
\label{o2n11}
\eea
The $\cN=1$ superfields $\eta_{-n+2}, \dots ,\eta_{-1}$ are complex unconstrained, 
while $\eta_0$ is real unconstrained.\footnote{In the special case $n=1$, 
which corresponds to the $\cN=2$ tensor multiplet \cite{KLR}, 
the component $\eta_0$ is a real linear $\cN=1$  superfield.} 
Finally, the components $\eta_1, \dots, \eta_n$
are related to those already considered by complex conjugation, eq. (\ref{o2n1}).
${}$For the other projective multiplets, $ \O_{a}^{(n_{a})} $, entering the Lagrangian  
$\cL^{(2)} (\eta^{(2n)},  \O_{a}^{(n_{a})}  ; v)$, 
we appropriately replace $ \O_{a}^{(n_{a})}  (v ) \to  \O_{a}^{[n_{a}]} (\z)$. 
The supersymmetric action turns into
\bea
S &=&  \oint_\g 
 \frac{\rd\z }{ 2\pi\ri  \z}
\int\rd^4 x\,{\rm d}^4\q \, 
\cL (\eta^{[2n]},  \O_{a}^{[n_{a}]}  ; \z) ~.
\eea

Now, let us turn to the dual formulation. In complete analogy with $\eta^{(2n)}$,
associated with the real tropical multiplet $U^{(2n)} (v) $ is the superfield $U^{[2n]}(\z) $ defined by 
(\ref{2n-tropica1}).
Associated with the arctic multiplet $\X^{(2-2n)} (v) $ is  $\X^{[2-2n]}(\z) $ defined by 
\bea
\X^{(2-2n)} (v) &=&\big({\rm i})^{1-n} (v^{\1} \big)^{2-2n} \X^{[2-2n]}(\z) ~, \non \\
\X^{[2-2n]}(\z) &=& 
\sum_{k=0}^{\infty} \X_k  \,\z^k~, \qquad 
{\bar D}_{\dt \a} \X_0 =0~, 
\qquad {\bar D}^2 \X_1 =0~ .
\eea
For the smile-conjugate antarctic multiplet, $\breve{\X}^{(2-2n)} (z,v) $, we get
 \bea
\breve{\X}^{(2-2n)} (v) &=&\big({\rm i})^{n-1} (v^{\1} \z \big)^{2-2n} \breve{\X}^{[2-2n]}(\z) ~, \quad
\breve{\X}^{[2-2n]}( \z) = \sum_{k=0}^{\infty}  {\bar \X}_k \,
\frac{(-1)^k}{\z^k}~.~~~
\eea

The first-order action becomes
\bea
S_{\text{first-order}}  &=& 
 \oint_\g   \frac{\rd\z }{ 2\pi\ri  \z}
\int\rd^4 x\,{\rm d}^4\q \,\Big\{ 
\cL (U^{[2n]},  \O_{a}^{[n_{a}]}  ; \z)  \non \\
&&{}\qquad \qquad \qquad + U^{[2n]} \Big( \z^{n-1} \X^{[2-2n]} + \big(-\z \big)^{1-n}\,
\breve{\X}^{[2-2n]} \Big)\Big\} ~.~~~
\eea
This action coincides in form with that introduced in \cite{LR} (see also \cite{G-RRWLvU}).

\subsection{Polar-polar duality}
\label{sub4.2}
We turn to  a $\cN=1$ formulation for the theory with Lagrangian 
 $\cL^{(2)} (\U^{(n)}, \breve{\U}^{(n)} ,\O ; v)$ and its dual version considered 
 in subsection \ref{Polar-polar duality I}.

The  arctic multiplet $\U^{(n)} ( v) $ is represented by the series (\ref{arctic1}),
in which the two leading components $\U_0$ and $\U_1$ are, respectively, 
chiral and complex linear $\cN=1$ superfields,
\bea
{\bar D}_{\dt \a} \U_0 =0~, 
\qquad {\bar D}^2 \U_1 =0~ ,
\eea
while the other components $\U_2, \U_3, \dots$, are complex unconstrained $\cN=1$ superfields.
Its smile-conjugate antarctic multiplet, $\breve{\U}^{(n)} (v) $, is given by eq. (\ref{antarctic1}).
${}$For the other projective multiplets, $ \O_{a}^{(n_{a})} $, in the Lagrangian  
 $\cL^{(2)} (\U^{(n)}, \breve{\U}^{(n)} , \O_{a}^{(n_{a})}  ; v)$,
we appropriately replace $ \O_{a}^{(n_{a})}  (z,v ) \to  \O_{a}^{[n_{a}]} (z,\z)$. 
The supersymmetric action turns into
\bea
S &=&  \oint_\g 
 \frac{\rd\z }{ 2\pi\ri  \z}
\int\rd^4 x\,{\rm d}^4\q \, 
 \cL (\U^{[n]}, \breve{\U}^{[n]} ,  \O_{a}^{[n_{a}]}  ; \z) ~.
\eea

Consider now  the dual formulation.
Associated with the complex tropical multiplet $W^{(n)} (z,v) $ 
is the superfield $W^{[n]}(z,\z) $ defined by 
\bea
W^{(n)} (v) &=&\big( v^{\1}  \big)^n W^{[n]}(\z) ~,  \qquad
W^{[n]}(\z) = 
\sum_{k=-\infty}^{\infty} W_k  \z^k~.
\eea
For its smile-conjugate antarctic multiplet, $\breve{W}^{(n)} (z,v) $, we get
\bea
\breve{W}^{(n)} (v) &=& \big(v^{\1} \z \big)^{n}\, \breve{W}^{[n]}(\z) ~, \qquad
\breve{W}^{[n]}( \z) = \sum_{k=-\infty}^{\infty}  {\bar W}_k \,
\frac{(-1)^k}{\z^k}~.~~~
\eea
Finally, the arctic  superfield ${\X}^{(2-n)} (z,v) $ and its smile-conjugate 
 $\breve{\X}^{(2-n)} (z,v) $ will be represented  similarly to eqs. 
 (\ref{arctic1}) and (\ref{antarctic1}) with the replacement $n \to 2-n$.
 The first-order action becomes
 \bea
S_{\text{first-order}} &=&  \oint_\g 
 \frac{\rd\z }{ 2\pi\ri  \z}
\int\rd^4 x\,{\rm d}^4\q \, \Big\{  
 \cL (W^{[n]}, \breve{W}^{[n]} , \O_{a}^{[n_{a}]}  ; \z) \non \\
&&{}\qquad \qquad \qquad  +\frac{1}{\z} W^{[n]} \X^{[2-n]}   
 -\z \breve{W}^{[n]}\breve{\X}^{[2-n]}   \Big\} ~.~~~~
\eea
This formulation of the polar-polar duality coincides with that given in \cite{GK,LR10}.
 
\subsection{Polar-polar duality and superconformal $\s$-models}
Of special interest for us is 
the superconformal $\s$-model defined by eqs. (\ref{conformal-sm}) and (\ref{Kkahler2}), 
for it can be argued to realize general $\cN=2$ superconformal couplings.
We represent the weight-one arctic multiplets as $\U^{+ I} (v)  =  v^{\1} \, \U^I ( \z)$, where 
\bea
\U^I ( \z) = \sum_{k=0}^{\infty} \U^I_k  \z^k=\F^I + \z \, \S^I  + O(\z^2) ~, 
\qquad
{\bar D}_{\dt \a} \F^I =0~, 
\quad {\bar D}^2 \S^I =0~ .~~~
\label{3.10}
\eea
We recall that the components $\U_2, \U_3, \dots$, are complex unconstrained 
$\cN=1$ superfields.
The $\cN=2$ superconformal action turns into
\bea
S &=&  \oint_\g 
 \frac{\rd\z }{ 2\pi\ri  \z}
\int\rd^4 x\,{\rm d}^4\q \, 
 K (\U^I , \breve{\U}^{\bar J} ) ~.
 \label{3.17}
\eea

The dual formulation is described by the following $\cN=2$ superconformal action:
\bea
S_{\rm D} &=&  \oint_\g 
 \frac{\rd\z }{ 2\pi\ri  \z}
\int\rd^4 x\,{\rm d}^4\q \, 
 K_{\rm D} (\X_I , \breve{\X}_{\bar J} ) ~.
\eea
Here the arctic multiplets  $\X_I (\z) $ is related to $\X^+_I(v)$ by the rule
 $\X^+_{ I} (v)  =  v^{\1} \, \X_I ( \z)$, and the structure of  $\X_I (\z) $ is completely similar to that given 
 in eq. (\ref{3.10}).
The dual Lagrangian is defined by 
\bea
 K_{\rm D} (\X_I , \breve{\X}_{\bar J} ) = 
 \Big\{  K (W, \breve{W} ) +\frac{1}{\z} W^I \X_I   
 -\z \breve{W}^{\bar I} \breve{\X}_{\bar I}   \Big\} \Big|~,
 \label{4.21}
 \eea 
where the tropical superfields $W^I$ and $\breve{W}^{\bar I}$ 
must be unique solutions of the algebraic equations:
\bea
 \frac{\pa}{\pa W^{I} } K (W, \breve{W}) &+& \frac{1}{\z}  \X_I =0~, \qquad
 \frac{\pa}{\pa \breve{W}^{\bar I} } K (W, \breve{W}) - \z \breve{\X}_{\bar I} =0~.
\label{4.22}
\eea

The K\"ahler potential $K (\F, {\bar \F} )  $ and its dual $K_{\rm D}(\J , {\bar \J} ) $ 
correspond, in general,  to different K\"ahler cones. As will be argued in the remainder of this
paper, both potentials are encoded in
the hyperk\"ahler potential,  ${\mathbb K} (\F, \bar \F, \J, \bar \J )$, 
in the target space for the original $\s$-model (\ref{3.17}). 
It will also be shown that the original hyperk\"ahler potential  
${\mathbb K} $ and its dual ${\mathbb K}_{\rm D} $ 
are related to each other by a holomorphic reparametrization.

To conclude this section, we summarize, without proof,  the explicit structure of
the hyperk\"ahler potential  ${\mathbb K} (\F^{I}, {\bar \F}^{\bar I}, \J_J, {\bar \J}_{\bar J} )$; 
the technical details can be found  in \cite{K-duality}. It has the form:
\bea
{\mathbb K} (\F, \bar \F, \J, \bar \J )
=K \big( \F, \bar{\F} \big)+  \cH(\F, \bar \F, \J, \bar \J )
\label{H-Kpotential}
\eea
where the second term  obeys the homogeneity condition 
\bea
\Big( \F^I \frac{\pa}{\pa \F^I}  +  { \J}_{ I}   \frac{\pa }{\pa { \J}_{ I} }
 \Big) \cH \big(\F, \bar \F, \J , \bar \J \big)
= \cH \big(\F, \bar \F, \J , \bar \J \big)~,
\label{homo5}
\eea
as well as the condition 
\bea
{\J}_{I} \frac{\pa \cH}{\pa {\J}_{ I} }  &=&
{\bar \J}_{\bar I}
\frac{\pa \cH}{\pa {\bar \J}_{\bar I} }  ~,
\label{semi-homo4} 
\eea
and hence 
\bea
{\F}^{I} \frac{\pa \cH}{\pa {\F}^{ I} }  &=&
{\bar \F}^{\bar I}
\frac{\pa \cH}{\pa {\bar \F}^{\bar I} }  ~.
\eea

\section{Polar-polar duality with a single hypermultiplet}
\setcounter{equation}{0}

In this section we carry out a more systematic study of the polar-polar duality.
In particular, we provide proofs for several statements, specifically 
eqs. (\ref{3.7}), (\ref{3.8a}) and (\ref{3.8b}), which were taken for granted 
in our previous consideration. For the sake of simplicity, our discussion is restricted 
to the case of $\cN=2$ supersymmetric sigma-models described by a single polar multiplet.
However, many results can be readily extended to 
the case of $n$ polar multiplets.  
All nontrivial $\cN=2$ supersymmetric sigma-models  of the type specified (that is, described 
by one polar multiplet) are non-superconformal, except 
$\cL = \breve{\U}\, \U $,  when $\U (\z)$  has weight one.

\subsection{General analysis}
\label{general}
Consider an off-shell $\cN=2$ supersymmetric nonlinear $\s$-model 
described by a polar multiplet realized in terms of 
an arctic superfield $\U (\z) $ and its 
smile conjugate $\breve{\U}( \z)$. 
\bea
S &=&  
 \oint_\g \frac{{\rm d}\z}{2\pi {\rm i} \z} \,  
 \int  {\rm d}^4 x \, {\rm d}^4\q\,  \cL \big( \U , \breve{\U}; \z   \big) ~,
\label{act-hyper} 
\eea
where $\U (\z)$ looks like 
\bea
 \U (\z) &=& \sum_{n=0}^{\infty}  \, \z^n \U_n  = 
\F + \z \, \S  + O(\z^2) ~,\qquad
{\bar D}_{\dt{\a}} \F =0~, \quad {\bar D}^2 \S = 0 ~,
\label{5.2new}
\eea
and its smile-conjugate $\breve{\U} (\z)$ has the form
\bea
\breve{\U} (\z) &=& \sum_{n=0}^{\infty}  \,  (-\z)^{-n}\,
{\bar \U}_n~.
\eea
We recall that the components $\U_2, \U_3, \dots$, in (\ref{5.2new}) are complex unconstrained
$\cN=1 $ superfields. These superfields appear in the action without derivatives, and therefore 
they are purely auxiliary.

Before discussing the dual formulation for the theory (\ref{act-hyper}), it is worth recalling 
the explicit structure of the equations of motion, see, e.g., \cite{GK,LR10}.
Since the $\cN=1$ superfields $\U_2, \U_3, \dots$, in (\ref{5.2new}) are complex unconstrained,
their equations of motion 
\bea
\frac{\d }{\d \U_n} S =0~, \qquad n \geq 2
\eea
have the form:
\bea
\oint_\g \frac{{\rm d} \z}{\z} \,\z^n \, \frac{\pa \cL(\U, \widetilde{\U};\z ) }{\pa \U} 
 = 0 ~, \qquad n \geq 2 ~ .              
\label{asfem-con}
\eea
This infinite set of nonlinear algebraic equations  are equivalent to 
\bea
\frac{\pa }{\pa \U} \cL \big( \U , \breve{\U}; \z   \big) +\frac{1}{\z} \P =0~, \qquad
 \P (\z) &:=& \sum_{n=0}^{\infty}  \, \z^n \P_n ~,
\label{asfem}
\eea
for some superfield $\P(\z)$. These equations can be used, in principle, to express
the auxiliary superfields $\U_n$, with $n\geq 2$,
in terms of the physical superfields $\F$ and $\S$ and their conjugates;
after that, the explicit form of $\P (\z)$ can be determined as well. 
It remains to consider the equations of motion for the physical
chiral ($\F:=\U_0$) and complex linear ($\S:=\U_1$) superfields.\footnote{In deriving the equations of motion 
for $\F$ and $\S$, it is useful to represent $\F = {\bar D}^2 \bar R$ and $\S= {\bar D}_{\dt \a} {\bar \x}^{\dt \a}$, 
for unconstrained superfields $\bar R$ and ${\bar \x}^{\dt \a}$.} These equations imply that 
$\P (\z)$ is an arctic multiplet. We see that the off-shell constraint ($\U (\z)$ is arctic) 
and the equation of motion ($\P(\z)$ is arctic) have the same superfield type.
This is characteristic of duality-covariant theories in which the Bianchi identities and equations 
of motion have the same functional type (see, e.g., \cite{HL} for a review).

Now, let us apply the polar-polar duality transformation to  the action (\ref{act-hyper}). Following
subsection \ref{sub4.2}, we replace (\ref{act-hyper}) by the  first-order action
\bea
S_{\text{first-order}} &=&  
 \oint_\g \frac{{\rm d}\z}{2\pi {\rm i} \z} \,  
 \int  {\rm d}^4 x \, {\rm d}^4\q\,  \Big\{ \cL \big( W , \breve{W}; \z   \big) 
 +\frac{1}{\z} W \X   -\z \breve{W}\breve{\X}   \Big\} ~,
\label{act-hyper-fo} 
\eea
where $W(\z)$ is  complex tropical,
\bea
 W (\z) &=& \sum_{n=-\infty}^{\infty}  \, \z^n W_n  ~, \qquad 
 \breve{W} (\z) = \sum_{n=-\infty}^{\infty}  \,  (-\z)^{-n}\,
{\bar W}_n  ~,
\label{5.8}
\eea
and the Lagrange multiplier $\X(\z)$ is arctic,
\bea
 \X (\z) &=& \sum_{n=0}^{\infty}  \, \z^n \X_n  = 
\J + \z \, \G  + O(\z^2) ~,\qquad
{\bar D}_{\dt{\a}} \J =0~, \quad {\bar D}^2 \G = 0 ~.
 \eea
We would like to show that 
the theory with action  (\ref{act-hyper-fo}) is equivalent to the original one, eq. (\ref{act-hyper}).
To vary (\ref{act-hyper-fo}) with respect to $\X(\z)$, it is useful 
{\it (i)} to do the contour  integrals in the second 
and third terms on the right of (\ref{act-hyper-fo}), as well as 
{\it (ii)} to separate  the contributions involving the auxiliary 
and the physical superfields contained in $\X(\z)$:
\bea
S_{\text{first-order}} &=&  
 \oint_\g \frac{{\rm d}\z}{2\pi {\rm i} \z} \,  
 \int  {\rm d}^4 x \, {\rm d}^4\q\,   \cL \big( W , \breve{W}; \z   \big) 
+\sum_{n=2}^{\infty}  \int  {\rm d}^4 x \, {\rm d}^4\q\,
\Big\{ \X_n  W_{-n+1} +{\rm c.c.}\Big\}\non \\
&&+  \int  {\rm d}^4 x \, {\rm d}^4\q\,
\Big\{ \J W_{1} + +\G W_0 + {\rm c.c.}\Big\}~.
\label{act-hyper-fo2} 
\eea
Since the superfields 
$\X_n$, with $n=2,3,\dots$, are complex unconstrained, their equations of motion are
\bea
W_{-n+1}&=&0~, \qquad n \geq 2~.
\label{15a}
\eea
Next, the equations of motion for $\J$ and $\G$ are equivalent to the conditions  
that $W_1$ and $W_0$ are complex linear and chiral, respectively.
Our conclusion is thus the following:
\bea 
\frac{\d }{\d \X} S_{\text{first-order}}=0 \quad \Longrightarrow \quad
W(\z) = \U(\z)~.
\label{5.12}
\eea 
As a result,
the second and third terms in  (\ref{act-hyper-fo2}) drop out, and
$S_{\text{first-order}} $ reduces to (\ref{act-hyper}).

The above derivation of eq. (\ref{5.12}) can be readily extended to justify  the equation   (\ref{3.7})
in the general case.

On the other hand, instead of varying  (\ref{act-hyper-fo})  with respect to $\X(\z)$, 
we can first vary $S_{\text{first-order}}$ with respect to $W(\z)$.
Since all the components $W_n$ in  (\ref{5.8}) are complex unconstrained superfields, 
we immediately obtain
\bea
\frac{\d }{\d W} S_{\text{first-order}}=0 \quad \Longrightarrow \quad
\frac{\pa }{\pa W} \cL \big( W , \breve{W}; \z   \big) +\frac{1}{\z} \X =0~.
\label{5.13}
\eea
This equation and its smile-conjugate can be used to 
express $W(\z)$ in terms of $\X(\z)$, $\breve{\X}(\z)$ and $\z$.
As a result, $S_{\text{first-order}} $ turns into the dual action
\bea
S_{\rm D} &=&  
 \oint_\g \frac{{\rm d}\z}{2\pi {\rm i} \z} \,  
 \int  {\rm d}^4 x \, {\rm d}^4\q\,  \cL_{\rm D} \big( \X , \breve{\X}; \z   \big) ~.
\label{act-hyper-dual} 
\eea

Our derivation of eq. (\ref{5.13}) can be readily generalized to justify  the equation (\ref{3.8a}).

\subsection{Chiral-linear duality}
\label{chiral-linear}
The $\cN=2$ supersymmetric nonlinear $\s$-model (\ref{act-hyper}) 
can be formulated solely in terms of the physical 
superfields $\F$, $\S$ and their conjugates.
The equations (\ref{asfem}) can be used to express all the auxiliary superfields $\U_2, \U_3, \dots$
(as well as the components $\P_n$ in  (\ref{asfem}))
in terms of the physical ones. Then, the action  (\ref{act-hyper}) turns into
the {\it chiral-linear} (CL) one
\bea
S^{\rm(CL)} &=&  
 \int  {\rm d}^4 x \, {\rm d}^4\q\,  L^{\rm(CL)}  \big( \F, \bar \F; \S, \bar \S   \big) ~.
\label{act-CL} 
\eea

The chiral-linear formulation can also be obtained for the dual theory (\ref{act-hyper-dual}) 
following the same rules. This leads to
\bea
S^{\rm(CL)}_{\rm D} &=&  
 \int  {\rm d}^4 x \, {\rm d}^4\q\,  L^{\rm(CL)}_{\rm D}  \big( \J, \bar \J; \G, \bar \G   \big) ~.
\label{act-CL-dual} 
\eea
We now demonstrate that $L^{\rm(CL)}_{\rm D}  \big( \J, \bar \J; \G, \bar \G   \big) $
is a Legendre transform of $ L^{\rm(CL)}  \big( \F, \bar \F; \S, \bar \S   \big) $.

Let us return to the first-order formulation (\ref{act-hyper-fo}) for the theory (\ref{act-hyper}). 
This  first-order action  is equivalent to (\ref{act-hyper-fo2}). 
Consider the equations of motion for the auxiliary superfields
$\X_n$ and $W_{n}$, where $n\geq 2$.
The equations of motion for $\X_n$ , with $n\geq 2$, are given by (\ref{15a}).
 The  equations of motion for $W_{n}$, with $n\geq 2$, are
\bea
\frac{\pa }{\pa W} \cL \big( W , \breve{W}; \z   \big) +\frac{1}{\z} \L &=&0~, \qquad
\L (\z) := \sum_{n=0}^{\infty}  \, \z^n \L_n ~,
\label{15b}
\eea
with $\L (\z)$ some superfield.
Eq. (\ref{15a}) tells us that $W(\z)$ is now represented by a Taylor series.
Eq. (\ref{15b}) has the same functional form as the auxiliary field equation of motion, 
eq. (\ref{asfem})), in the theory (\ref{act-hyper}). 
Therefore, making use of eqs. (\ref{15a}) and (\ref{15b})  allows us to transform (\ref{act-hyper-fo2}) 
to the form:
\bea
S'_{\text{first-order}} &=& 
 \int  {\rm d}^4 x \, {\rm d}^4\q\,  L^{\rm(CL)}  \big( U, \bar U; V, \bar V  \big) 
 + \int  {\rm d}^4 x \, {\rm d}^4\q\,
\Big\{ \G U - \J V+ {\rm c.c.}\Big\}~,~~~
\label{act-hyper-fo3} 
\eea
where we have denoted
\bea
U:= W_0~, \qquad V:= -W_1~.
\eea
It is clear that  the first-order model (\ref{act-hyper-fo3}) is equivalent to  (\ref{act-hyper-fo}).
The latter is also equivalent to (\ref{act-CL-dual})
This indeed shows that $L^{\rm(CL)}_{\rm D}  \big( \J, \bar \J; \G, \bar \G   \big) $
is a Legendre transform of $ L^{\rm(CL)}  \big( \F, \bar \F; \S, \bar \S   \big) $.

As is clear from the above consideration, 
the transformation
$$ L^{\rm(CL)}  \big( \F, \bar \F; \S, \bar \S   \big) \quad \longrightarrow \quad
L^{\rm(CL)}_{\rm D}  \big( \J, \bar \J; \G, \bar \G   \big) $$
actually involves two independent Legendre transformations: 

(a) dualization of 
the (anti) chiral variables $\bar \F$ and $\F$ into (anti) linear ones 
$\bar \G $ and $\G$;  

(b) dualization of 
the (anti) linear variables $\bar \S$ and $\S$ into (anti) chiral ones 
$\bar \J $ and $\J$.

\noindent
It is easy to see the order in which these Legendre transformations is performed 
(say, first carry out (a) and then (b), or vise versa)
does not matter. We can also apply  single Legendre transformations, 
specifically:
\bea
 L^{\rm(CL)}  \big( \F, \bar \F; \S, \bar \S   \big) \quad & \longrightarrow &\quad
L^{\rm(CC)}  \big( \F, \bar \F; \J, \bar \J   \big) ~,\\
L^{\rm(CL)}  \big( \F, \bar \F; \S, \bar \S   \big) \quad & \longrightarrow &\quad
L^{\rm(LL)}  \big( \G, \bar \G; \S, \bar \S   \big) ~.
\eea
The Lagrangians obtained
can be further Legendre-transformed
\bea
L^{\rm(CC)}  \big( \F, \bar \F; \J, \bar \J   \big)\quad & \longrightarrow &\quad
L^{\rm(LC)}  \big( \G, \bar \G; \J, \bar \J   \big) ~, \\
L^{\rm(LL)}  \big( \G, \bar \G; \S, \bar \S   \big)
\quad & \longrightarrow &\quad
L^{\rm(LC)}  \big( \G, \bar \G; \J, \bar \J   \big) ~,
\eea
where the notation introduced should be quite transparent. 
One has
\bea
L^{\rm(LC)}  \big( \G, \bar \G; \J, \bar \J   \big) 
= L^{\rm(CL)}_{\rm D}  \big( \J, \bar \J; \G, \bar \G   \big) ~.
\label{5.24}
\eea

It should be pointed out that 
\bea
{\mathbb K} (\F, \bar \F, \J, \bar \J ):= L^{\rm(CC)}  \big( \F, \bar \F; \J, \bar \J   \big)
\eea
coincides with the hyperk\"ahler potential in the target space of the 
$\cN=2$ supersymmetric  $\s$-model (\ref{act-hyper}). 
Let ${\mathbb K}_{\rm D} (\J, \bar \J, \F, \bar \F )$ be the 
 hyperk\"ahler potential in the target space of the dual model
(\ref{act-hyper-dual}). It follows from (\ref{5.24}) that
\bea
{\mathbb K}_{\rm D} (\J, \bar \J, \F, \bar \F ) ={\mathbb K} (\F, \bar \F, \J, \bar \J )~.
\eea

\subsection{$\cN=2$ $\s$-models on  cotangent bundles of K\"ahler manifolds}
Let us now consider  those  $\s$-models (\ref{act-hyper}) in which the Lagrangian 
has no  explicit dependence  on $\z$, 
\bea
\cL \big( \U , \breve{\U}; \z   \big)  \quad \longrightarrow \quad 
\cL \big( \U , \breve{\U}\big)
=  K \big( \U , \breve{\U}   \big) ~.
 \label{5.16}
\eea
Here $K(\F , \bar \F)$ is real analytic function that can be consistently interpreted  \cite{GK} 
as the K\"ahler potential of a two-dimensional K\"ahler manifold $\cM$.
The action (\ref{act-hyper}) associated with the Lagrangian (\ref{5.16})
is invariant under U(1) transformations of the form:
\bea
\U(\z) ~ \longrightarrow ~ \U'(\z) =
\U( {\rm e}^{{\rm i}\b} \z)~, \qquad \b \in \mathbb R~.
\label{5.17}
\eea
This invariance follows from the fact that the contour integration measure in (\ref{act-hyper})
is invariant under transformations $\z \to {\rm e}^{{\rm i}\b} \z$,\footnote{Transformations 
$ \z \to {\rm e}^{{\rm i} \b} \zeta$ can be interpreted as 
 time translations along $\g$. This becomes manifest if  the integration  contour 
 $\g$  in  (\ref{act-hyper}) is chosen to be  $\z(t)= R \,{\rm e}^{{\rm i} t}$. Thus, 
 if $\z$ is viewed as a complex evolution parameter,  the Lagrangian
 (\ref{5.16}) is a generalization of mechanical systems 
 with conserved energy (for such a system,   its Lagrangian  $L(q, {\dt q} )$ has no
 explicit time dependence).} 
 and thus
 \bea
 \oint_\g \frac{{\rm d}\z}{2\pi {\rm i} \z} \,  
  \cL \Big( \U' (\z), \breve{\U}' (\z) ; \z   \Big) 
= \oint_\g \frac{{\rm d}\z}{2\pi {\rm i} \z} \,  
   \cL \Big( \U (\z), \breve{\U} (\z) ; {\rm e}^{-{\rm i} \b} \z   \Big) ~,
\eea
with $\U'(\z) $ defined in (\ref{5.17}).
The hyperk\"ahler potential in  the target space turns out to have the form:
\bea
{\mathbb K} (\F, \bar \F, \J, \bar \J )
=K \big( \F, \bar{\F} \big)+  \cH(\F, \bar \F, \J, \bar \J )~,
\label{HK-potential}
\eea
where the complex variables $(\F, \J)$ parametrize (an open domain of the zero section of) 
the holomorphic cotangent bundle $T^*\cM$ of the K\"ahler manifold $\cM$.\footnote{More generally, 
in the case of $n$ self-interacting polar multiplets described by a $\z$-independent Lagrangian,
$\cL =  K \big( \U^I , \breve{\U}^{\bar J}   \big) $, it was shown in \cite{GK,GK2}
that the $\s$-model target space is (an open domain of the zero section of) 
the cotangent bundle $T^*\cM$ of a K\"ahler manifold $\cM$ for which $K \big( \F^I , \bar{\F}^{\bar J}   \big) $
is the K\"ahler potential. This supersymmetric $\s$-model result implies the existence of a hyperk\"ahler structure
on $T^*\cM$, for an arbitrary real-analytic K\"ahler manifold \cite{GK,GK2}. The latter result
 was independently  established by purely mathematical means \cite{Kaledin,Feix}.}
The second term in (\ref{HK-potential})
must be invariant under arbitrary phase transformations of the one-form $\J$,
\bea
\cH(\F, \bar \F,{\rm e}^{{\rm i}\b} \J, {\rm e}^{-{\rm i}\b}\bar \J )=\cH(\F, \bar \F, \J, \bar \J )~.
 \qquad \b \in \mathbb R~.
\eea

The hyperk\"ahler potential must obey the Monge-Amp\`ere equation 
(see, e.g., \cite{Besse})
\bea
\det \, \left(\begin{array}{cc}
 \frac{\pa^2 \mathbb K}{\pa \F \pa {\bar \F}} 
 ~ &   \frac{\pa^2 \mathbb K}{\pa \F \pa {\bar \J} }  \\
 \frac{\pa^2 \mathbb K}{\pa \J \pa {\bar \F} } 
~ &  \frac{\pa^2 \mathbb K}{\pa \J \pa {\bar \J} } 
\end{array}
\right)=1~.
\label{MA}
\eea
It can be represented in the form:
\bea
\cH(\F, \bar \F, \J, \bar \J ) =\sum_{n=1}^{\infty} 
H_n(\F, \bar \F ) 
\left[  \frac{\J \,\bar \J}{g_{\F \bar \F} (\F, \bar \F)}   \right]^n~, \qquad H_1 =1
\label{H-series}
\eea
where $g_{\F \bar \F} (\F, \bar \F) =\pa_{\F} \pa_{\bar \F} K \big( \F, \bar{\F} \big)$ 
is the K\"ahler metric on $\cM$, and $H_n(\F, \bar \F ) $ are real analytic {\it scalar} fields on $\cM$.
We would like to rewrite the Monge-Amp\`ere equation in terms of objects intrinsic to the K\"ahler base
manifold. From the point of view of the K\"ahler base, $\cH$ can be thought of as a linear combination
of tensor fields where $\J$ play the role of base co-vectors. By the definition of the covariant derivative
on the K\"ahler base manifold we realize that
\bea
\nabla_\F \cH = \sum_{n=1}^{\infty}
\nabla_\F  H_n (\F, \bar \F ) 
\left[ \frac{\J \,\bar \J}{g_{\F \bar \F} (\F, \bar \F)} \right]^n~.
\eea
Using this, the Monge-Amp\`ere equation is equivalent to 
\bea
g_{\F \bar \F} \frac{\pa^2 \cH}{\pa \J \pa {\bar \J} } - 1&=&
\Big(\nabla_\F \frac{\pa \cH}{ \pa {\bar \J} }\Big) \nabla_{\bar \F} \frac{\pa \cH}{\pa \J  } \non \\
&&-\Big[ \nabla_{\bar \F} \nabla_\F \cH
+\hf g_{\F \bar \F}\,R\, 
\J \frac{\pa \cH}{\pa \J} 
\Big]\frac{\pa^2 \cH}{\pa \J \pa {\bar \J} }~,
\label{MA2}
\eea
with $R$ denoting the scalar curvature of $\cM$, that is $ g_{\F \bar \F}\,R= -2 
\pa_{\bar \F} \G^\F_{\F \F}$. In the right-hand side of (\ref{MA2}), the covariant derivatives $\nabla_{ \F} $ and 
$\nabla_{\bar \F} $ act only on the scalar fields $H_n$ appearing in (\ref{H-series}),
that is 
\bea
( \nabla_{\bar \F})^a  (\nabla_\F)^b \cH (\F, \bar \F, \J, \bar \J ) :=
\sum_{n=1}^{\infty} 
\Big[ ( \nabla_{\bar \F})^a  (\nabla_\F)^bH_n(\F, \bar \F ) \Big]
\Big[  \frac{\J \,\bar \J}{g_{\F \bar \F} (\F, \bar \F)}   \Big]^n~.
\eea
Eq. (\ref{MA2}) is equivalent to a recursion relation  to 
uniquely compute the coefficients $H_n$.
The recursion relation is as follows:
\bea
H_m &=& \frac{1}{m^2}\sum_{n=1}^{m-1} \frac{m-n}{g_{\F \bar \F} }
\Big(n
( \nabla_\phi H_n )\nabla_{\bar\phi}H_{m-n}
-(m-n)(\nabla_\phi\nabla_{\bar\phi} H_n )H_{m-n}
\nonumber\\
&&
\qquad \qquad
-\frac{1}{2}n(m-n) g_{\F \bar \F}  R H_n H_{m-n}\Big)~,\qquad m\geq 2~.
\eea

To construct the dual formulation,  we should consider the first-order action
\bea
S_{\text{first-order}} &=&  
 \oint_\g \frac{{\rm d}\z}{2\pi {\rm i} \z} \,  
 \int  {\rm d}^4 x \, {\rm d}^4\q\,  \Big\{ K \big( W , \breve{W}   \big) 
 +\frac{1}{\z} W \X   -\z \breve{W}\breve{\X}   \Big\} ~, 
\label{5.29}
\eea
with $W(\z)$  a complex tropical multiplet, and $\X(\z)$ an  arctic superfield.
The U(1) symmetry of the original model,  eq. (\ref{5.17}), turns into
\bea
W(\z) ~ \longrightarrow ~ W({\rm e}^{{\rm i}\b} \z)~, \qquad 
\X(\z) ~ \longrightarrow ~ {\rm e}^{-{\rm i}\b} \X( {\rm e}^{{\rm i}\b} \z)~.
\eea
From (\ref{5.29}) we read off the dual Lagrangian
\bea
\cL_{\rm D} \big( \X , \breve{\X}; \z   \big)  =  -K_{\rm D} \big(- \z^{-1} \X , \z \breve{\X}   \big) 
\label{5.37}
\eea
for some real function $K_{\rm D} (\J, \bar \J)$. 
Unlike the original Lagrangian, eq. (\ref{5.16}), 
the dual Lagrangian depends, in general,  on $\z$. Such a dependence disappears 
in special cases which will be discussed below.

\subsection{$\cN=2$ $\s$-models with U(1)$\times$U(1)  symmetry}
\label{sub5.4}

Here we would like to consider a subclass of hypermultiplet models (\ref{5.16})
which are invariant under two rigid U(1) symmetries: phase transformations
\bea
\U(\z) ~ \longrightarrow ~ {\rm e}^{{\rm i}\a} \U(\z)~, \qquad \a \in \mathbb R
\label{5.38}
\eea
and shadow chiral rotations\footnote{Such transformations naturally originate 
in $\cN=2$ superspace parametrized by $z^A =(x^a, \q^\a_i, {\bar \q}^i_{\dt \a})$, 
with $i =\1, \2$, 
as part of the $R$-symmetry group ${\rm SU(2) \times U(1)}$, see \cite{K-hyper} for more
details. A  shadow chiral rotation  is a phase transformation
of $\q^\a_{\2}$ only, with $\q^\a_{\1}$ kept unchanged.
The $\cN=1$ superspace is identified with  the surface $\q^\a_{\2}=0$.}
\bea
\U(\z) ~ \longrightarrow ~ \U'(\z) ={\rm e}^{-({\rm i}/2)\b}\U({\rm e}^{{\rm i}\b} \z)~, \qquad \b \in \mathbb R~.
\label{5.39}
\eea
The most general Lagrangian compatible with such symmetries is
\bea
 \cL \big( \U , \breve{\U}; \z   \big)  =  \cL \big( \U  \breve{\U}   \big) ~,
 \label{5.43}
\eea
with $\cL (x) $ a real analytic function of one real variable. We can interpret 
$K(\F, \bar \F):= \cL (\F \bar \F)$ as the K\"ahler potential of a two-dimensional space $\cM$
in {\it canonical} (or {\it K\"ahler normal}) complex  coordinates, see subsection \ref{sub6.1}.

With the above Lagrangian, the first-order action (\ref{act-hyper-fo}) takes the form
\bea
S_{\text{first-order}} &=&  
 \oint_\g \frac{{\rm d}\z}{2\pi {\rm i} \z} \,  
 \int  {\rm d}^4 x \, {\rm d}^4\q\,  \Big\{ \cL \big( W  \breve{W}   \big) 
 +\frac{1}{\z} W \X   -\z \breve{W}\breve{\X}   \Big\} ~, 
\eea
with $W(\z)$  a complex tropical multiplet, and $\X(\z)$ an  arctic multiplet.
This action is invariant  under arbitrary phase transformations
\bea
W(\z) ~ \longrightarrow ~ {\rm e}^{{\rm i}\a} W(\z)~, \qquad 
\X(\z) \quad \longrightarrow \quad {\rm e}^{-{\rm i}\a} \X(\z)
\eea
and shadow chiral rotations
\bea
W(\z) ~ \longrightarrow ~ {\rm e}^{-({\rm i}/2)\b}W({\rm e}^{{\rm i}\b} \z)~, \qquad 
\X(\z) ~ \longrightarrow ~ {\rm e}^{-({\rm i}/2)\b} \X( {\rm e}^{{\rm i}\b} \z)~.
\eea
These symmetries are therefore present in the dual theory (\ref{act-hyper-dual}). 
As a result,  the corresponding Lagrangian  has the form
\bea
\cL_{\rm D} \big( \X , \breve{\X}; \z   \big) = \cL_{\rm D} \big( \X \, \breve{\X}   \big) ~.
\eea
The dual Lagrangian, $\cL_{\rm D} $,  and the original one, $\cL $,
are related to each other as follows
\bea
\cL_{\rm D} \big( \X \, \breve{\X}   \big) =  \cL \big( W  \breve{W}   \big) 
-2  \cL' \big( W  \breve{W}   \big) \,W  \breve{W}~,
\eea
where $W$ and its smile-conjugate $\breve W $ are to be expressed via
$\X$ and $ \breve \X$ using the equations
\bea
\cL' \big( W  \breve{W}   \big) \,  \breve{W}  +\frac{1}{\z}\,  \X=0~, \qquad 
\cL' \big( W  \breve{W}   \big) \,  {W} - {\z} \, \breve{\X}=0~.
\eea 
 
The dual Lagrangian $\cL_{\rm D} $ can be given a geometric interpretation
of the K\"ahler potential, $K_{\rm D}(\J, \bar \J):= \cL_{\rm D} (\J \bar \J)$, 
of a two-dimensional space $\cM$ with ${\rm U(1) \times U(1)}$ isometry.
By construction, the K\"ahler potential is given
in {\it canonical}  complex  coordinates.
 
It turns out that both the original   K\"ahler potential $K(\F, \bar \F)$ and its dual
$K_{\rm D}(\J, \bar \J)$ are  encoded in the hyperk\"ahler potential 
(\ref{HK-potential}).  Before turning to a detailed justification of this claim, it is worth considering an example.
 
\subsection{The Eguchi-Hanson metric and polar-polar duality}
\label{EH}
As an instructive example of the sigma-models studied in the previous subsection,
consider the K\"ahler potential corresponding to ${\mathbb C}P^1$:
\bea
K(\F, \bar \F) = \ln \,(1 + \F \, \bar \F )~.
\label{EH1}
\eea
Associated with this potential is the polar multiplet Lagrangian:
\bea
 \cL \big( \U , \breve{\U}   \big) =  \ln\,\big(1+  \U \, \breve{\U}   \big) ~.
\label{EH-polar}
\eea
A short calculation gives for the dual Lagrangian:
\bea
 \cL_{\rm D} \big( \X , \breve{\X}   \big) =  \sqrt{ 1+  4\U \, \breve{\U} } -1
 - \ln \frac{1+ \sqrt{1+  4\U \, \breve{\U}}  }{2} ~.
 \label{EH-polar2}
\eea
The dual Lagrangian is associated with the K\"ahler potential 
 \bea
K_{\rm D} \big( \J , \bar{\J}   \big) =  \sqrt{ 1+  4\J \, \bar{\J} } -1
 - \ln \frac{1+ \sqrt{1+  4\J \, \bar{\J}}}{2} ~.
\label{EH2}
\eea
which corresponds to a new K\"ahler manifold that differs from the two-sphere.
This follows from the fact that the K\"ahler metric for ${\mathbb C}P^1$, 
\bea
g_{\F \bar \F} = (1 + \F \, \bar \F )^{-2}~, 
\eea
is characterized by a constant curvature,  while the K\"ahler metric generated by 
the dual K\"ahler potential (\ref{EH2})), 
\bea
g_{\J \bar \J} = (1 + \J \, \bar \J )^{-1/2}~, 
\eea
is {\it no longer a metric of constant curvature}.  In addition, the dual K\"ahler manifold is non-compact, 
unlike ${\mathbb C}P^1$.
 
To get a better understanding of the relationship between the K\"ahler potential (\ref{EH1}) 
and its dual (\ref{EH2}), consider the hyperk\"ahler potential generated by the $\cN=2$ 
supersymmetric Lagrangian (\ref{EH-polar}).\footnote{Within the projective superspace approach,
the hyperk\"ahler potential (\ref{5.53}) was computed in \cite{GK,AKL}.} 
It is 
\bea
{\mathbb K} ( \F, \bar{\F} , \J , \bar \J ) &=&
K(\F, \bar \F) +   \sqrt{1+4|\J|^2 } -1 -\ln \frac{1+ \sqrt{1+4|\J|^2} }{2} ~, 
\label{5.53} \\
|\J|^2 &:=& 
\frac{\J \, \bar \J}{ g_{\F \bar \F}}  = (1 + \F \, \bar \F )^2 \J \, \bar \J~. \non
\eea
The K\"ahler potential (\ref{EH1}) 
and its dual (\ref{EH2}) can be seen to correspond to two different limits one can define 
in terms of the hyperk\"ahler potential ${\mathbb K} ( \F, \bar{\F} , \J , \bar \J ) $:
\begin{subequations}
\bea
{\mathbb K} ( \F, \bar{\F} , 0 , 0 ) &=& K(\F, \bar \F)~, \\
{\mathbb K} ( 0, 0 , \J , \bar \J ) &=&K_{\rm D} \big( \J , \bar{\J}   \big)~.
\eea
\end{subequations}
Since these limits are defined in terms of local complex coordinates on $T^*{\mathbb C}P^1$,
one might think they are non-geometric. This is not quite true, for the coordinate system used in 
(\ref{5.53}) is canonical, and canonical coordinates  for K\"ahler manifolds
\cite{Bochner} are intrinsic (they are defined modulo  linear holomorphic reparametrizations), 
see subsection \ref{sub6.1}.

\section{Polar-polar duality with $ n$ hypermultiplets}
\setcounter{equation}{0}
This section generalizes the analysis given in subsections \ref{sub5.4} and \ref{EH}
to the case of $n$ interacting hypermultiplets.

\subsection{Canonical coordinates for K\"ahler manifolds}
\label{sub6.1}
We start by recalling the concept of canonical coordinates for K\"ahler manifolds
\cite{Bochner}.
Given a K\"ahler manifold $\cM$,  
for any point $ p_0 \in \cM$ there exists  a  neighborhood of $p_0$ such that 
holomorphic reparametrizations  and K\"ahler transformations
can be used to choose  coordinates with origin at $p_0$ in which
the K\"ahler potential is
\bea
{K} (\F, \bar \F ) &=&{g}_{I \bar{J}}| \,\F^I {\bar \F}^{\bar J}
+ \sum^{\infty}_{ m,n \geq 2}  
{ { K}^{(m,n)} (\F, \bar \F)}~,
\non \\
 { K}^{(m,n)} (\F, \bar \F) &:=&
\frac{1}{m! n!}\,
{ K}_{I_1 \cdots I_m {\bar J}_1 \cdots {\bar J}_n }|  \, \F^{I_1} \dots \F^{I_m} 
{\bar \F}^{ {\bar J}_1 } \dots {\bar \F}^{ {\bar J}_n }~.
\label{normal-gauge} 
\eea
Such a coordinate system in the K\"ahler manifold is called {\it canonical}. 
It was first introduced by Bochner \cite{Bochner} and extensively used by Calabi 
in the 1950s \cite{Calabi}.\footnote{In the  modern literature on supersymmetric $\s$-models, 
some authors, unaware of the work of \cite{Bochner},
refer to the canonical coordinates as a 
{\it normal gauge} \cite{GGRS} or {\it K\"ahler normal coordinates} \cite{HIN}.}
There still remains  freedom to perform 
linear holomorphic reparametrizations which can be used 
to set the metric at the origin, $p_0 \in \cM$, to  be ${ g}_{I \bar{J}}|= \d_{I \bar{J}}$.
The resulting frame is defined modulo  linear holomorphic U($n$) transformations.

It turns out that the  coefficients ${ K}_{I_1 \cdots I_m {\bar J}_1 \cdots {\bar J}_n } |$
in (\ref{normal-gauge})
are tensor functions of the K\"ahler metric ${g}_{I \bar{J}}|$,
the Riemann curvature $R_{I {\bar J} K {\bar L}}  |$ and its covariant 
derivatives, all evaluated at the origin. In particular, one finds
\begin{subequations}
\bea
{K}^{(2,2)} &=& \frac{1}{4} R_{I_1 {\bar J}_1 I_2 {\bar J}_2} |\, \F^{I_1}\F^{I_2}
{\bar \F}^{{\bar J}_1}{\bar \F}^{{\bar J}_2}~,
\label{(2,2)}  \\
 {K}^{(3,2)} &=& \frac{1}{12} 
\nabla_{I_3} R_{I_1 {\bar J}_1 I_2 {\bar J}_2}| \,
\F^{I_1} \dots \F^{I_3}
{\bar \F}^{{\bar J}_1}{\bar \F}^{{\bar J}_2}~, 
\label{(3,2)} \\
 {K}^{(4,2)} &=& \frac{1}{48} 
\nabla_{I_3} \nabla_{I_4} R_{I_1 {\bar J}_1 I_2 {\bar J}_2} |\,
\F^{I_1} \dots \F^{I_4}
{\bar \F}^{{\bar J}_1}{\bar \F}^{{\bar J}_2}~, 
\label{(4,2) } \\
{K}^{(3,3)} &=& \frac{1}{12} \Big\{ \frac{1}{6} 
\{ \nabla_{I_3}, {\bar \nabla}_{{\bar J}_3} \}
R_{I_1 {\bar J}_1 I_2 {\bar J}_2} |
+R_{I_1 {\bar J}_1 I_2 }{}^L |R_{L {\bar J}_2 I_3 {\bar J}_3}| \Big\}  \non \\
 &&\quad  \times \F^{I_1} \dots \F^{I_3}
{\bar \F}^{{\bar J}_1}\dots{\bar \F}^{{\bar J}_3}~~~~~~~
\label{(3,3)}  
\eea
\end{subequations}
The functions ${K}^{(4,3)}$ and ${K}^{(4,4)}$ are given in \cite{K-hyper}.

\subsection{$\cN=2$ $\s$-models with U(1)$\times$U(1)  symmetry}
Here we consider a family of K\"ahler manifolds $\cM$ with holomorphic U(1) isometry.
In canonical coordinates, the corresponding K\"ahler potential  is
\bea
{K} (\F^I, {\bar \F}^{\bar J} ) &=&{g}_{I \bar{J}}| \,\F^I {\bar \F}^{\bar J}
+ \sum^{\infty}_{ n \geq 2}  
{ { K}^{(n,n)} (\F, \bar \F)}~,
\non \\
 { K}^{(n,n)} (\F, \bar \F) &:=&
\frac{1}{(n!)^2}\,
{ K}_{I_1 \cdots I_n {\bar J}_1 \cdots {\bar J}_n }|  \, \F^{I_1} \dots \F^{I_n} 
{\bar \F}^{ {\bar J}_1 } \dots {\bar \F}^{ {\bar J}_n }~.
\label{normal-gauge2} 
\eea
The relevant isometry acts as a phase transformation
\bea
\F^I ~ \longrightarrow ~ {\rm e}^{{\rm i}\a} \F^I~, \qquad \a \in \mathbb R
\eea
which leaves the K\"ahler potential invariant.
The condition (\ref{normal-gauge2}) is equivalent to 
\bea
\F^I \frac{\pa}{\pa \F^I} K(\F, \bar \F) = {\bar \F}^{\bar J} \frac{\pa}{\pa {\bar \F}^{\bar J}} K(\F, \bar \F)~.
\label{normal-gauge22} 
\eea
All Hermitian symmetric spaces belong to this family of manifolds.
Note that eq. (\ref{normal-gauge22})  is weaker than 
the homogeneity condition (\ref{Kkahler2})
which corresponds to the superconformal action (\ref{3.17}).  

Associated with such a K\"ahler manifold $\cM$ is the $\cN=2$ 
supersymmetric $\s$-model
\bea
S &=&  \oint_\g 
 \frac{\rd\z }{ 2\pi\ri  \z}
\int\rd^4 x\,{\rm d}^4\q \, 
 K (\U^I , \breve{\U}^{\bar J} ) ~.
 \label{6.5}
\eea
This model possesses ${\rm U(1)} \times {\rm U(1)}$  symmetry. 
The relevant symmetry transformations are derived from 
(\ref{5.38}) and (\ref{5.39})  simply by replacing $\U (\z) \to \U^I(\z)$.
The model (\ref{6.5}) is non-superconformal except for the trivial case of a quadratic K\"ahler potential.

We apply the polar-polar duality transformation to all the arctic multiplets $\U^I$ 
and their conjugates in (\ref{6.5}).
The dual Lagrangian $ K_{\rm D} (\X_I , \breve{\X}_{\bar J} ) $, which is defined 
as in eqs.  (\ref{4.21}) and (\ref{4.22}), 
has the same functional form as the K\"ahler potential in 
eq. (\ref{normal-gauge2}). We conclude that polar-polar duality generates a
transformations between K\"ahler spaces, $\cM \to \cM_{\rm D}$,  described (in canonical coordinates)
by  K\"ahler potentials of the form (\ref{normal-gauge2}).

There is a simple relationship between the K\"ahler potential  $K (\F^I, {\bar \F}^{\bar J} ) $ and 
its dual $ K_{\rm D} (\J_I , \bar{\J}_{\bar J} )$. Let us first discuss the structure 
of the hyperk\"ahler target space for the $\cN=2$ supersymmetric $\s$-model (\ref{6.5}). 
As argued in \cite{GK,GK2}, the $\s$-model target space is an open domain of the zero
section of the cotangent bundle $T^*\cM$ parametrized by complex variables
$(\F^I, \J_I)$ and their conjugates, with $\J_I$ a holomorphic one-form 
at the point $(\F, \bar \F)$ of the base space $\cM$. The hyperk\"ahler potential 
for $T^*\cM$ can be chosen as
\bea
{\mathbb K} (\F, \bar \F, \J, \bar \J ) =
K(\F, \bar \F) +  \cH(\F, \bar \F, \J , \bar \J )~,
\label{bold-K}
\eea
where the function $\cH(\F, \bar \F, \J, \bar \J )$   can be represented by a Taylor series 
\bea
\cH \big(\F, \bar \F, \J , \bar \J \big)&=& 
\sum_{n=1}^{\infty} \cH^{I_1 \cdots I_n {\bar J}_1 \cdots {\bar 
J}_n }  \big( \F, \bar{\F} \big) \J_{I_1} \dots \J_{I_n} 
{\bar \J}_{ {\bar J}_1 } \dots {\bar \J}_{ {\bar J}_n } ~,\non \\
\cH^{I {\bar J}} \big( \F, \bar{\F} \big) 
&=& g^{I {\bar J}} \big( \F, \bar{\F} \big) ~.
\label{h}
\eea
Here the Taylor coefficients $\cH^{I_1 \cdots I_n {\bar J}_1 \cdots {\bar J}_n }  \big( \F, \bar{\F} \big) $
are some tensor functions of the K\"ahler metric
$g_{I \bar{J}} \big( \F, \bar{\F}  \big) 
= \pa_I 
\pa_ {\bar J}K ( \F , \bar{\F} )$, 
 the Riemann curvature $R_{I {\bar 
J} K {\bar L}} \big( \F, \bar{\F} \big) $ and its covariant 
derivatives.  One can see from (\ref{bold-K}) that the complex coordinate system 
$(\F^I, \J_I)$ in $T^*\cM$  is {\it canonical}.

Next, let us turn to the dual $\s$-model. Its target space is an open domain of the zero
section of the cotangent bundle $T^*\cM_{\rm D}$ parametrized by complex variables
$(\J_I, \F^I)$ and their conjugates. The hyperk\"ahler potential 
for $T^*\cM_{\rm D}$ is
\bea
{\mathbb K}_{\rm D} (\J, \bar \J, \F, \bar \F) =
K_{\rm D}(\J, \bar \J) +  \cH_{\rm D}(\J , \bar \J, \F, \bar \F)~,
\label{bold-K2}
\eea
where $ \cH_{\rm D}$ has a series representation which is similar to that given in (\ref{h}) 
and is obtained from the latter by the replacement $\F \longleftrightarrow \J$
(including the replacement of all relevant geometric objects).

${}$Finally, we can apply the chiral-linear duality of  subsection \ref{chiral-linear} to show that 
\bea
{\mathbb K}_{\rm D} (\J, \bar \J, \F, \bar \F) = {\mathbb K} \big(\F, \bar \F, \J , \bar \J \big)~.
\label{6.10}
\eea
We can immediately conclude that 
\bea
{\mathbb K} ( \F, \bar{\F} , 0 , 0 ) &=& K(\F, \bar \F)~, \qquad
{\mathbb K} ( 0, 0 , \J , \bar \J ) =K_{\rm D} \big( \J , \bar{\J}   \big)~.
\label{6.11}
\eea
As a result, we see that the K\"ahler potential  $K (\F, {\bar \F} ) $ and 
its dual $ K_{\rm D} (\J , \bar{\J} )$ are encoded in the hyperk\"ahler potential 
$ {\mathbb K} \big(\F, \bar \F, \J , \bar \J \big)$.

\subsection{$\cN=2$ $\s$-models on  cotangent bundles of Hermitian symmetric spaces}
Hermitian symmetric spaces form a subclass in the family of K\"ahler manifolds
introduced in the previous subsection.
If $\cM$ is Hermitian symmetric, then  
\bea
\nabla_L  R_{I_1 {\bar  J}_1 I_2 {\bar J}_2}
= {\bar \nabla}_{\bar L} R_{I_1 {\bar  J}_1 I_2 {\bar J}_2} =0
\qquad \Longrightarrow \qquad 
{ K}^{(m,n)}=0~, \quad m\neq n~.
\label{covar-const}
\eea
This follows from the fact that, 
for Hermitian symmetric spaces, there exists a closed-form expression for
the K\"ahler potential in the canonical coordinates \cite{KN}:
\bea
K \big( \F, \bar \F   \big) = - \hf {\bm \F}^{\rm T} {\bm g} \,
\frac{ \ln \big( {\mathbbm 1} - {\bm R}_{\F,\bar \F}\big)}{\bm R_{\F,\bar \F}}
\, {\bm \F}~, \qquad 
{\bm \F} :=\left(
\begin{array}{c}
\F^I\\
{\bar \F}^{\bar I} 
\end{array}
\right) ~.
\eea
Here we have introduced
\bea
{\bm g}
&:=&\left(
\begin{array}{cc}
0 & g_{I \bar J}|\\
g_{{\bar I}J} |&0 
\end{array}
\right)
~, \qquad
{\bm R}_{\F,\bar \F}
:= \left(
\begin{array}{cc}
0 & (R_\F)^I{}_{\bar J}\\
(R_{\bar \F})^{\bar I}{}_J &0 
\end{array}
\right)~,  \non \\
(R_\F)^I{}_{\bar J} &:=&\hf R_K{}^I{}_{L \bar J}|\, \F^K \F^L~,  
\qquad (R_{\bar \F})^{\bar I}{}_J := \overline{(R_\F)^I{}_{\bar J}}~.
\eea

The program of deriving the hyperk\"ahler potential on $T^*\cM$ 
from the $\s$-model  (\ref{6.5}), for various Hermitian symmetric spaces,
has been carried in a series of papers \cite{GK,GK2,AN,AKL,AKL2}.
It has resulted in a universal expression for the hyperk\"ahler potential 
derived in \cite{KN} using the results of  \cite{AKL2}.
It is given by eq. (\ref{bold-K}),
where
\bea
\cH(\F, \bar \F, \J , \bar \J ) = \hf {\bm \J}^{\rm T}{\bm g}^{-1}  \cF \Big( - {\bm R}_{\J,\bar \J} \Big)\, 
{\bm \J} ~, \qquad 
{\bm \J} :=\left(
\begin{array}{c}
\J_I\\
{\bar \J}_{\bar I} 
\end{array}
\right) ~.
\label{hyperkahler-potential}
\eea
Here the function $\cF(x)$ is defined as
\bea
 \cF(x) &:=& 
 \frac{1}{x} \,\Big\{ \sqrt{1+4x} -1 -\ln \frac{1+ \sqrt{1+4x} }{2} \Big\}~, 
\qquad \cF(0)=1~,~~~~~~
\label{F-Calabi}
\eea
and the operator $ {\bm R}_{\J,\bar \J} $ has the form:
\bea
{\bm R}_{\J,\bar \J}
&:=&\left(
\begin{array}{cc}
0 & (R_\J)_I{}^{\bar J}\\
(R_{\bar \J})_{\bar I}{}^J &0 
\end{array}
\right)~, \non \\ 
(R_\J)_I{}^{\bar J}&=& (R_\J)_{IK} \,g^{K \bar J}~, \qquad 
(R_\J )_{K L}:= \hf R_{K}{}^I{}_{L}{}^J \,\J_I \J_J~. 
\eea
In the canonical coordinates, the curvature $R_{K}{}^I{}_{L}{}^J$ is a constant tensor, 
see, e.g., \cite{KN} for more details.

Using eq. (\ref{6.11}), from (\ref{hyperkahler-potential}) we can immediately 
read off the polar-polar dual of the $\s$-model (\ref{6.5}) if 
$ K (\F , \bar{\F}) $ is the K\"ahler potential of a Hermitian symmetric space.

\section{Self-dual hypermultiplet models}
\setcounter{equation}{0}

The concept of polar-polar duality allows us to introduce self-dual hypermultiplet models.
${}$For simplicity, here we consider the case of a single polar hypermultiplet.

The theory with action (\ref{act-hyper}) is said to be self-dual if the dual Lagrangian coincides
with the original one, 
\bea
\cL_{\rm D} \big( \U , \breve{\U}; \z   \big) = \cL \big( \U , \breve{\U}; \z   \big) ~.
\label{self-dual}
\eea
The simplest example of self-dual systems was given in \cite{GK}. 
It is the free hypermultiplet model
\bea
\cL_{\rm free} (\U, \breve{\U})= \U \breve{\U}~.
\eea
Below we construct an infinite family of self-dual nonlinear hypermultiplet 
models.

\subsection{The meaning of self-duality}

If the off-shell $\s$-model is self-dual, eq. (\ref{self-dual}), then we also have
\bea
L^{\rm(CL)}_{\rm D}  \big( \F, \bar \F; \S, \bar \S   \big) 
= L^{\rm(CL)} \big( \F, \bar \F; \S, \bar \S   \big) ~.
\label{self-dual2}
\eea
This is equivalent to the condition 
\bea
L^{\rm(CL)} \big( \J, \bar \J; \G, \bar \G   \big)
= L^{\rm(CL)}  \big( U, \bar U; V, \bar V  \big) 
+\G U + \bar \G \bar U~-~ 
 \J V -\bar \J \bar V~,
\label{self-dual3}  
\eea
where $U$ and $V$ are functions of $\J$, $\G$ and their conjugates 
which have to be determined by solving  the equations
\begin{subequations}
\bea
\frac{\pa}{\pa U} L^{\rm(CL)} \big( U, \bar U; V, \bar V   \big)+\G &=&0~, \\ 
\frac{\pa}{\pa V} L^{\rm(CL)} \big( U, \bar U; V, \bar V   \big)-\J &=&0~.
\eea
\end{subequations}

In the case of a self-dual model, one can readily see that 
the Lagrangains $L^{\rm(CC)}  \big( \F, \bar \F; \J, \bar \J   \big)$ and 
$L^{\rm(LL)}  \big( \G, \bar \G; \S, \bar \S   \big) $ must be symmetric functions.
\bea
L^{\rm(CC)}  \big( \F, \bar \F; \J, \bar \J   \big)
&=& L^{\rm(CC)}  \big( \J, \bar \J; \F, \bar \F   \big)~,
\label{symmetry1} \\
L^{\rm(LL)}  \big( \G, \bar \G; \S, \bar \S   \big)&=& L^{\rm(LL)}  \big( \S, \bar \S; \G, \bar \G   \big)
~.
\label{symmetry2}
\eea

In accordance with the above consideration, 
the nonlinear $\s$-model
\bea
S=  \int  {\rm d}^4 x \, {\rm d}^4\q\,L^{\rm(CC)}  \big( \F, \bar \F; \J, \bar \J   \big)
\eea
is $\cN=2$ supersymmetric, and therefore $L^{\rm(CC)}  \big( \F, \bar \F; \J, \bar \J   \big)$
should be the hyperk\"ahler potential of a hyperk\"ahler manifold $\cM$ \cite{HKLR}.
If the original off-shell action (\ref{act-hyper}) is self-dual, then 
the hyperk\"ahler potential $L^{\rm(CC)}  \big( \F, \bar \F; \J, \bar \J   \big)$
must be symmetric, eq. (\ref{symmetry1}). 
This means that the target space $\cM$ must possess a ${\mathbb Z}_2$ symmetry.

\subsection{Self-duality equation}
In this section we consider a simple special class of self-dual models. We assume
that the dependence on the polar multiplet is through the combination
$x = \Upsilon\breve\Upsilon$ with no explicit $\zeta$-dependence so that the
Lagrangian is given by an ordinary function $\cL(x)$.

Suppose the  theory under consideration is self-dual, $\cL_{\rm D}  =\cL$.
Then the Lagrangian $\cL(x) $ can be seen to obey the algebraic equation
\begin{subequations}
\bea
\cL' (x) \, \cL' (y) =1~, \qquad 
\label{sde1}
\eea
where the variables $x$ and $y$ are related to each other as follows
\bea
y = - x\, [\cL' (x)]^2 ~.
\label{sde2}
\eea
It follows from eqs. (\ref{sde1}) and (\ref{sde2}) that 
\bea
x = - y\, [\cL' (y)]^2 ~.
\label{sde3}
\eea
\end{subequations}

In particular, any function $f(x)$ with the property that $x=f(f(x))$ gives a self-dual
Lagrangian through
\bea
\cL(x) = \int^x {\rm d}x^{\prime}\, \sqrt{-\frac{f(x^\prime)}{x^\prime}}~.
\eea
Self-dual Lagrangians are not rare. Given any function $h(x)$ we can
construct a function satisfying the above properties as $f(x) = h^{-1}(-h(x))$
(compare with \cite{G-RKPR}).
If the function $h(x)$ is even or odd, the solution is trivial $f(x) = \pm x$, so
for nontrivial solutions we need functions $h(-x)\neq \pm h(x)$.

A simple partial solution of the equations (\ref{sde1}) and (\ref{sde2})  is
\bea
\cL(x) = - \frac{2}{g^2}\Big\{ 1- \sqrt{1+g^2 x} \Big\}~,
\eea
with $g$ a coupling constant.

To work out the geometry of this self-dual model one could use that models
with this particular dependence on the polar multiplet can be dualized to an
$\cO (2)$ multiplet $\eta$ as
\bea
\cL_2(\eta) =-\frac{2}{g^2} + \eta + \sqrt{\eta^2+\frac{4}{g^4}}
-\eta\ln\left(\frac{g^2\eta^2}{2}+\sqrt{\eta^2+\frac{g^4\eta^4}{4}}\right)~,
\eea
although there may appear contour ambiguities \cite{LR10}.
After working out the component content of the dual model one would have to
dualize the $\cN=1$ linear superfield of the $\cO(2)$ multiplet to get the
hyperk\"ahler potential.
We leave the explicit solution of this model for a future 
publication.

In theories with more polar multiplets there are other ways to construct self-dual
models. For instance if we have two polar multiplets we may start with an
action $\cL(\Upsilon_1\breve\Upsilon_1+\Upsilon_2\breve\Upsilon_2)$. Making
a polar-polar duality on only one of the polar multiplets we create a model which
will be self-dual with respect to a duality transformation of all polar multiplets.

\section{Final comments}
\setcounter{equation}{0}

In this paper, we demonstrated that polar-polar duality generates a transformation 
between different K\"ahler cones. This is a new type of duality.  It relates K\"ahler manifolds
that are target space geometries for $\cN=1$  $\s$-models, although 
the duality is intrinsically $\cN=2$ supersymmetric.
These dual K\"ahler manifolds are both embedded in the hyperk\"ahler target space 
of the original $\cN=2$ supersymmetric $\s$-model.

${}$For non-superconformal $\s$-models, we derived a simple relationship between 
the hyperk\"ahler potential ${\mathbb K}$ and its dual ${\mathbb K}_{\rm D}$, 
eq. (\ref{6.10}). It can naturally be extended to the superconformal case.

In order to find a hyperk\"ahler potential corresponding 
to a $\cN=2$ supersymmetric $\s$-model formulated in terms of polar multiplets,
one has to eliminate the auxiliary $\cN=1$ superfields. This technical problem has been  solved 
for the $\s$-models on cotangent bundles of Hermitian symmetric spaces.
The full solution is given in \cite{AKL2,KN}. 
Polar-polar duality allows us to extend the class of $\s$-models for which this is possible.
As an example, we looked at the Eguchi-Hanson geometry
whose known solution allowed us to solve the auxiliary field problem for the dual model
(\ref{EH-polar2}). The duality thus allows for a treatment of more complicated geometries.

We have discussed a family of self-dual models and identified some interesting features. 
The intrinsic meaning of self-duality as well as the relation between the geometry of the dual models 
remain to be understood, however. Here an example with the geometry worked out would be of great help.

${}$For more than one polar multiplet, 
there are more possibilities to construct self-dual models.
In particular, we can also consider self-dual superconformal models. 
Their properties remain to be investigated.

\bigskip

\noindent
{\bf Acknowledgements:}\\
We are happy to thank the organizers of the 30th Winter School ``Geometry and Physics'' at Srni, 
Czech Republic, where this project was initiated.
SMK and RvU are grateful to the Department of Physics and Astronomy at
Uppsala University for hospitality at final stages of this project.
The work of SMK and UL was supported in part by the Australian Research Council.
SMK also acknowledges partial support from  the Australian Academy of Science.
The research of UL was supported by VR grant 621-2009-4066.
The research of RvU was supported by a Czech Ministry of Education grant No.
MSM0021622409.

Projective superspace and its various implications have been frequently discussed with M. Ro\v cek over the years.
His inspiration and insights are gratefully acknowledged.

\appendix
%%%%%%%%%%%%%%%%%%%%%%%%%%%%%%%%%%%%%%%%%%%%%%%%%%
%%%%%%%%%%%%%%%%%%%%%%%%%%%%%%%%%%%%%%%%%%%%%%%%%%

\section{Superconformal Killing vectors}
\setcounter{equation}{0}
\label{Killing}

In this appendix we recall salient properties of the $\cN=2$ superconformal
Killing vectors, following \cite{Park,KT}.
 
In  $\cN=2$ superspace ${\mathbb R}^{4|8}$ parametrized  
by  coordinates  $ z^A = (x^a,  \q^\a_i, {\bar \q}^i_\dt{\a} )$,  with $i=\1,\2$,
a first-order differential operator
\be
\x = {\overline \x} =\x^A (z)D_A
= \x^a (z) \,\pa_a + \x^\a_i (z)\,D^i_\a
+ {\bar \x}_{\dt \a}^i (z)\, {\bar D}^{\dt \a}_i
\ee   
is called a $\cN=2$ superconformal  Killing vector
if it  obeys the condition 
\be
[\x \;,\; {\bar D}_i^\ad] \; \propto \; {\bar D}_j^\bd ~,
\label{4Dmaster0}
\ee   
which implies
\be
{\bar D}_i^{\dt \a }\,\x^\b_j =0~, \qquad
{\bar D}_i^{\dt \a } \x^{\dt \b \b} = 4{\rm i} \, \ve^{\dt \a{}\dt \b} \,\x^\b_i~.
\label{4Dmaster}
\ee
A short calculation gives 
\be
[\x \;,\; D^i_\a ] = - (D^i_\a \x^\b_j) D^j_\b
= { \o}_\a{}^\b  D^i_\b - 
\bar{ {\s}} \, D^i_\a
- {\L}_j{}^i \; D^j_\a ~,
\label{4DmasterN=2} 
\ee
where the parameters of  Lorentz (${\o}$) 
and scale-chiral (${\s}$) transformations are
\be
{\o}_{\a \b}(z) = -\frac{1}{2}\;D^i_{(\a} \x_{\b)i}\;,
\qquad {\s} (z) = \frac{1}{4}
{\bar D}^{\dt \a}_i {\bar \x}_{\dt \a}^{ i} ~.
\label{lor,weylN=2}
\ee
These parameters can be seen  to be chiral
\be
{\bar D}^{\dt \a}_{ i} {\o}_{\a \b}=0~,
\qquad {\bar D}^{\dt \a}_{ i} {\s} =0~.
\ee
The parameters ${\L}_j{}^i$ defined by
\bea
{\L}_j{}^i (z) &=& \hf \Big(D^i_\a \x^\a_j - \hf \d^i_j D^k_\a \x^\a_k \Big)
=-\hf \Big({\bar D}^{\dt \a}_j {\bar \x}^i_{\dt \a} 
-\hf \d^i_j {\bar D}^{\dt \a}_k {\bar \x}^k_{\dt \a} \Big)
~,\non \\
&&\qquad 
\L^{ij}=\L^{ji}~,
\qquad \overline{\L^{ij} } = \L_{ij}  
\label{lambdaN=2}
\eea
correspond to   SU(2) transformations.
One can readily check the identity 
\be
D^k_\a {\L}_j{}^i = -2 \Big( \d^k_j D^i_\a 
-\frac{1}{2} \d^i_j D^k_\a  \Big) {\s}~,
\label{an1N=2}
\ee
and therefore 
\bea
D^{(i}_\a \L^{jk)} = {\bar D}^{(i}_{\dt \a} \L^{jk)} =0~.
\label{L-an}
\eea
Comparing this with eq. (\ref{2.28}), we see that $\L^{(2)}(v):= \L_{ij} v^i v^j$ 
is an $\cO (2)$ multiplet.

\section{Tensor multiplet formulation for $\cN=2$ $\sigma$-models with  U(1)$\times$U(1) symmetry}
\setcounter{equation}{0}

In subsection \ref{sub5.4}, we discussed the polar-polar duality of $\cN=2$ $\s$-models 
(\ref{5.43}). There is a different dual formulation for such theories which is given in terms of 
 an ${\cal O}(2)$ multiplet
$\eta = \frac{\bar\varphi}{\zeta} +G-\varphi\zeta$ (also known as an $\cN=2$ tensor multiplet)
and a $\cN=2$  Lagrangian $\cL_2 (\eta )$. 
Here we will elaborate on the structure of such $\s$-models.\footnote{Four-dimensional quaternion  
K\"ahler metrics with torus symmetry were studied in \cite{CP}.}
We will assume that the contour
integral in the corresponding action has been done,
\bea
S &=&  \oint
 \frac{\rd\z }{ 2\pi\ri  \z}
\int\rd^4 x\,{\rm d}^4\q \, 
\cL_2 (\eta) = \int\rd^4 x\,{\rm d}^4\q \,H(\varphi  \bar\varphi,G)~,
\eea
where $H(\varphi  \bar\varphi,G)$  is the resulting $\cN=1$ Lagrangian.

In \cite{Lindstrom:1983rt} it is shown that the Lagrangian $H(\varphi \bar\varphi,G)$ 
must satisfy the Laplace equation
$\partial_\varphi\partial_{\bar\varphi} H + \partial^2_G H = 0$. This can be used
to find the form of $H(\varphi \bar\varphi,G)$ from the knowledge of only part of it. For
instance, using the ansatz 
\bea
H = \sum_{n=0}^{\infty} H_n(G) (\varphi\bar\varphi)^n~,
\eea
where
$H_0(G)$ is the $\varphi$ independent part of the Lagrangian, the Laplace
equation gives us the recursion relation
\bea
H_n^{\prime\prime}(G) = - (n+1)^2 H_{n+1}(G)~,
\eea
which can be solved given the initial data $H_0(G)$. The solution is
\bea
H_n = \frac{(-1)^n}{(n!)^2} \frac{{\rm d}^{2n}}{{\rm d}G^{2n}} H_0(G)~.
\eea
Then the full Lagrangian can be written compactly as
\bea
H = J_0\left(\sqrt{4\varphi\bar\varphi}\frac{\rm d}{{\rm d}G}\right) H_0(G)~,
\eea
where $J_0$ is a Bessel function.
The program can be tested on the $\cN=2$ improved tensor multiplet model
where $H_0 = -G\ln G$. 
Applying the differential operator defined above one indeed gets
\bea
\sqrt{G^2+4\varphi\bar\varphi}-G\ln\frac{G+\sqrt{G^2+4\varphi\bar\varphi}}
{2}~,
\eea
which agrees (up to terms annihilated by the superspace measure) with the
Lagrangian given in  \cite{Lindstrom:1983rt}.

One may also use this scheme if one knows instead the dependence on $\varphi$
but not on $G$. Making the ansatz
\bea
H = \sum_n \tilde{H}_n(4\varphi\bar\varphi)G^n~,
\eea
leads via the Laplace equation to the recursion relation
\bea
\tilde{H}_{n+2} = -4\frac{4\varphi\bar\varphi\tilde{H}_n^{\prime\prime}
+\tilde{H}_n^\prime}{(n+1)(n+2)}~.
\eea
We see that the recursion relation does not mix $H_n$ with odd and even $n$.
It means that to find the most general solution it is not enough to know $H_0$
but we also need to know $H_1$.
Thus the general solution is
\begin{subequations}
\bea
\tilde{H}_{2n}(x) &=& \frac{(-4)^n}{(2n)!}\Big(\frac{\rm d}{{\rm d}x}x\frac{\rm d}{{\rm d}x}\Big)^n
\tilde{H}_0(x)~,
\\
\tilde{H}_{2n+1} &=& \frac{(-4)^n}{(2n)!}\Big(\frac{\rm d}{{\rm d}x}x\frac{\rm d}{{\rm d}x}\Big)^n
\tilde{H}_1(x)~,
\eea
\end{subequations}
where $x = 4\varphi\bar\varphi$.
Then the full Lagrangian can be written as follows:
\bea
H = \cos\left(\sqrt{4G\frac{\rm d}{{\rm d}x}x\frac{\rm d}{{\rm d}x}}\right)
\left(\tilde{H}_0(x)+G\tilde{H}_1(x)\right)~.
\eea
Applying this to the improved tensor multiplet with $\tilde{H}_0 = \sqrt{x}$ and
$\tilde{H}_1 = \frac{1}{2}\ln\frac{x}{4}$ we find the correct solution. It is interesting
that since $G\tilde{H}_1 $ is in fact a solution to the Laplace equation in itself, the higher
odd powers are missing.

\small{

}

\end{document}